\documentclass[]{aastex631}

\usepackage{amsmath,amssymb}
\usepackage{isotope}
\usepackage{graphicx}

\submitjournal{ApJ}

\shorttitle{}
\shortauthors{Benhar et al.}

\begin{document}

\title{Modeling Neutron Star Matter in the Age of 
 Multimessenger Astrophysics}

\correspondingauthor{}
\email{}

\author{Omar Benhar}
\affiliation{INFN, Sezione di Roma, I-00185 Rome, Italy}
\affiliation{Department of Physics, Sapienza University, I-00185 Rome, Italy}

\author{Alessandro Lovato}
\affiliation{INFN-TIFPA, Trento Institute of Fundamental Physics and Applications, I-38123 Trento, Italy}
\affiliation{Physics Division, Argonne National Laboratory. Argonne, Illinois 60439, USA}

\author{Giovanni Camelio}
\affiliation{Nicolaus Copernicus Astronomical Center, Polish Academy of Science, Bartycka 18, 00-716 Warsaw, Poland}

\begin{abstract}
The interpretation of the available and forthcoming data obtained from 
multimessenger astrophysical observations\textemdash potentially providing unprecedented 
access to neutron star properties\textemdash will require the development of novel, accurate theoretical 
models of dense matter. Of great importance, in this context, will be the capability to devise a description 
of thermal effects applicable to the study of quantities other than the equation of state, such as the transport coefficients and
the neutrino mean free path in the nuclear medium.
The formalism based on correlated basis states and the cluster expansion technique has been previously employed to
derive a well-behaved effective interaction\textemdash suitable for use in standard perturbation theory\textemdash from a state-of-the-art nuclear 
Hamiltonian, including phenomenological two- and three-nucleon potentials.
Here, we provide a comprehensive and self-contained account of the extension of this approach to the treatment of finite-temperature effects, and
report the results of numerical calculations of a number of properties of nuclear matter with arbitrary neutron excess
and temperature up to 50~MeV. 
\end{abstract}


\section{Introduction} 
\label{sec:intro}

The first detection of a gravitational wave signal consistent with emission from a coalescing binary neutron-star 
system~\citep{PhysRevLett.119.161101}, supplemented  by the later observation of electromagnetic radiation 
by space- and ground-based telescopes~\citep{Abbott_2017}, arguably opened up a new age for
both astrophysics and nuclear physics research. 

In years to come, multimessenger observations are expected to provide
unprecedented information on neutron star structure and dynamics, which will allow to shed light
not  only on bulk properties of nuclear matter\textemdash such as the Equation of State (EOS) determining the neutron-star mass and 
radius~\citep[see, e.g.,][]{bauswein}\textemdash but also on
the underlying nuclear dynamics. The possibility to exploit the available data to infer direct
information on nucleon interactions at microscopic level has been recently analysed in the pioneering study of~\citet{bayes1}. 

To meet the challenges posed by the interpretation of upcoming data, theoretical models must be 
capable to provide a consistent description of both equilibrium and dynamical properties of neutron-star matter
in the temperature regime corresponding to $T\ll~m_\pi$\textemdash $m_\pi \approx 140$ MeV being the pion mass\textemdash where 
nucleons are believed  to be the 
dominant degrees of freedom. In addition to the EOS, 
these include the transport coefficients describing  matter
viscosity~\citep{bv,Giovanni:2022_1,Giovanni:2022_2} and heat transfer~\citep{PhysRevC.81.024305}, as well as the neutrino mean free path in the nuclear medium~\citep{LLB,LBGL}, 
which play a critical role in the evolution of proto neutron 
stars~\citep{gct}, as well as in the post-merger phase of neutron-star coalescence~\citep{figuraetal2020,figuraetal2021}. 
Owing to the complexity and non perturbative nature of nuclear forces, however, the achievement of the above goal involves 
non trivial conceptual and computational issues.

The equation of state of nuclear matter can be obtained from highly accurate {\it ab initio} calculations, performed using 
phenomenological Hamiltonians\textemdash strongly constrained by the observed properties of two- and three-nucleon 
systems\textemdash  
and advanced theoretical approaches for the solution of the quantum-mechanical many-body problem~\citep[for a recent review, see, e.g.,][]{NMT}. On the other hand, the present development of computational techniques does not 
allow to accurately describe transport phenomena or neutrino reaction rates using the same Hamiltonians. 

In nuclear many-body theory, the problem of the occurrence of non perturbative interactions is circumvented through a renormalisation of the bare Hamiltonian, leading to its 
replacement with a density-dependent {\it effective} Hamiltonian suitable for use in perturbation theory.
This scheme has been followed to derive effective interactions within the $G$-matrix approach~\citep[for a review, see][]{Baldo} or using the formalism of
Correlated Basis Functions and the cluster expansion technique~\citep{CLARK197989,OCS}. The resulting potentials are well behaved, and 
have been used to perform calculations of a variety of nuclear matter properties relevant to astrophysical processes~\citep[see][]{mif,akmal:1998}.
More recently, it has been suggested that effective interactions suitable for perturbative calculations can also be obtained combining potentials derived within 
chiral effective field theory ($\chi$EFT) and renormalisation group evolution to low momentum,~\citep[see, e.g.,][]{SRG3,Keller:2020qhx,Drischler:2021kxf}. As clearly stated by~\citet{Tews_ApJ}, 
$\chi$EFT appears to be 
inherently inadequate for applications in the high-density region relevant to neutron stars; see also~\citep{Benhar:IJMPE}. 
However, studies of the convergence of the chiral expansion up to about twice nuclear-saturation density
provide a quantitative estimate of the theoretical uncertainty associated with the derivation of the nuclear Hamiltonian. 

It has to be pointed out that the effective interactions obtained from realistic dynamical models 
are conceptually different from those designed to merely explain the empirical information on average properties of atomic nuclei and isospin-symmetric nuclear matter near equilibrium density, such as the Skyrme interactions employed, e.g., in the studies of~\citet{skyrme2} and~\citet{skyrme1}. Because they lack a connection with microscopic dynamics, these models are in fact unable to describe nucleon-nucleon scattering in the nuclear medium, and their applications to the study of neutron star properties are severely limited~\citep{PhysRevC.81.024305}.

Starting from the early 2000s, the Correlated Basis Function, or CBF, formalism has been 
exploited to derive effective interactions from phenomenological Hamiltonians  
comprising both two- and three- nucleon potentials~\citep{bv,Ale:DD,LLB}. The capability to embody into a density-dependent NN potential  the effects of irreducible three-nucleon interactions\textemdash which are known to become large, or even dominant, at high densities\textemdash is a critical feature of the CBF approach. The effective interaction obtained from the 
latest and most advanced implementation of this approach, 
thoroughly described by~\citet{LLB}, has been applied to the determination of a variety of properties of nuclear matter, at
both zero and nonzero temperature, within a unified scheme \citep{eos0,gct}. 

This article is aimed at providing a comprehensive and self-contained account of the extension of the work of~\citet{eos0} to the case of non vanishing temperature, 
including a detailed derivation of the procedure employed to achieve thermodynamic consistency. 
The conceptual analogy between the CBF effective interaction, obtained from renormalisation in coordinate space, and the 
low-momentum interactions obtained from renormalisation group evolution is also discussed.

The body of the paper is organised as follows. The dynamical model underlying nuclear many-body theory
and the derivation of the microscopic effective interaction using the CBF formalism 
and the cluster expansion technique are outlined in Section~\ref{sec:dynamics}.  Section~\ref{finite:T} is devoted to a 
discussion of the issues associated with the use of many-body perturbation theory at nonzero temperature, while 
the results of numerical calculations of selected properties of hot nuclear matter\textemdash including the EOS, the 
single nucleon spectra, and the effective masses\textemdash and neutron stars 
are reported in Sections~\ref{EOS} and~\ref{nstar}, respectively.
Finally, our main findings and the prospects for future developments of our work are summarised in Section~\ref{summary}.

\section{Nuclear Dynamics} 
\label{sec:dynamics}
Ideally, a model of nuclear dynamics should be capable to provide a unified description of all
nucleon systems, from the deuteron to neutron stars~\citep{bob_rmp,NMT}.
In this section, we give an overview of the prominent issues associated with the treatment  of nuclear 
interactions at supranuclear densities, and outline the key elements of the approach employed in our work.
\subsection{The nuclear Hamiltonian} 
\label{sec:ham}
Nuclear many-body theory is founded on the hypothesis that nuclear systems can be treated 
as collections of 
point-like protons and neutrons, whose dynamics are
dictated by a non relativistic Hamiltonian of the form\footnote{Throughout the paper we will adopt the system of
natural units, in which $\hbar = c = k_B = 1$, and neglect the small proton-neutron mass difference.}

\begin{align}
H = \sum_{i=1}^{A} \frac{{\bf p}_i^2}{2m} + \sum_{j>i=1}^{A} v_{ij}
 + \sum_{k>j>i=1}^A V_{ijk} \ .
\label{H:A}
\end{align}
In the above equation, $A$ is the number of nucleons, ${\bf p}_i$ and $m$ denote the momentum of the $i$-th nucleon and its 
mass, and the potentials
$v_{ij}$ and $V_{ijk}$ describe two- and three-nucleon interactions, respectively.

The nucleon-nucleon (NN) potential is designed  
to reproduce the
measured properties of the two-nucleon system, in both bound and scattering
states, and reduces to the Yukawa one-pion exchange (OPE) potential at large distances.
It can be conveniently written in the form
\begin{align}
v_{ij}=\sum_{p} v^{p}(r_{ij}) O^{p}_{ij} \ ,
\label{eq:NN_1}
\end{align}
where the functions $v^p$ only depend on the distance between the interacting particles, $r_{ij} = |{\bf r}_i - {\bf r}_j|$, while the operators 
$O^{p}_{ij}$ account for the strong spin-isospin dependence of nuclear forces, as well as for the occurrence of non-central interactions. 
The most important contributions to the sum appearing in the right-hand side of Eq.~\eqref{eq:NN_1} are those associated with the six operators
\begin{align}
O^{p \leq 6}_{ij} = [1, (\boldsymbol{\sigma}_{i}\cdot\boldsymbol{\sigma}_{j}), S_{ij}]
\otimes[1,(\boldsymbol{\tau}_{i}\cdot\boldsymbol{\tau}_{j})]  \ ,
\label{av18:2}
\end{align}
where $\boldsymbol{\sigma}_{i}$ and $\boldsymbol{\tau}_{i}$ are Pauli matrices acting in spin and isospin space, respectively, and the tensor 
operator $S_{ij}$ is defined as
\begin{align}
S_{ij}=\frac{3}{r_{ij}^2}
(\boldsymbol{\sigma}_{i}\cdot{\bf r}_{ij}) (\boldsymbol{\sigma}_{j}\cdot{\bf r}_{ij})
 - (\boldsymbol{\sigma}_{i}\cdot\boldsymbol{\sigma}_{j}) \ .
 \label{S12}
\end{align}
Note that the OPE potential can be written in terms of the $O^{p \leq 6}_{ij}$ defined by  Eqs.\eqref{av18:2} and \eqref{S12}.

State-of-the-art phenomenological models of $v_{ij}$, such as the Argonne $v_{18}$ (AV18) potential~\citep{Wiringa:1994wb}\textemdash providing an accurate fit of the NN scattering phase shifts, 
the low-energy NN scattering parameters, and deuteron properties
\textemdash include twelve additional terms. The operators corresponding to $p~=~7,\ldots,14$ are associated with non-static components of the potential,  notably spin-orbit terms, while those corresponding to $p=15,\ldots,18$ take into account small violations of charge symmetry
and charge independence.

The results discussed in this article have been obtained using the Argonne $v_{6}^\prime$ (AV6P) potential, constructed projecting the full AV18 on the basis of the six operators of Eqs.~\eqref{av18:2}-\eqref{S12}~\citep{Wiringa:2002ja}.
This potential  
reproduces the deuteron binding energy and electric quadrupole moment with accuracy of 1\%, and 4\%, respectively, and provides an excellent fit
of the phase shifts in the $^1{\rm S}_0$ channel\footnote{We use spectroscopic notation, according to which 
the two-nucleon state $^1{\rm S}_0$ corresponds to orbital angular momentum $\ell=0$, spin $S=0$ and total angular momentum $J=0$. The total isospin of this state, dictated by the 
requirement of antisymmetry under particle exchange, is $T=1$.}.

\begin{figure}[htb]
\includegraphics[scale=0.65]{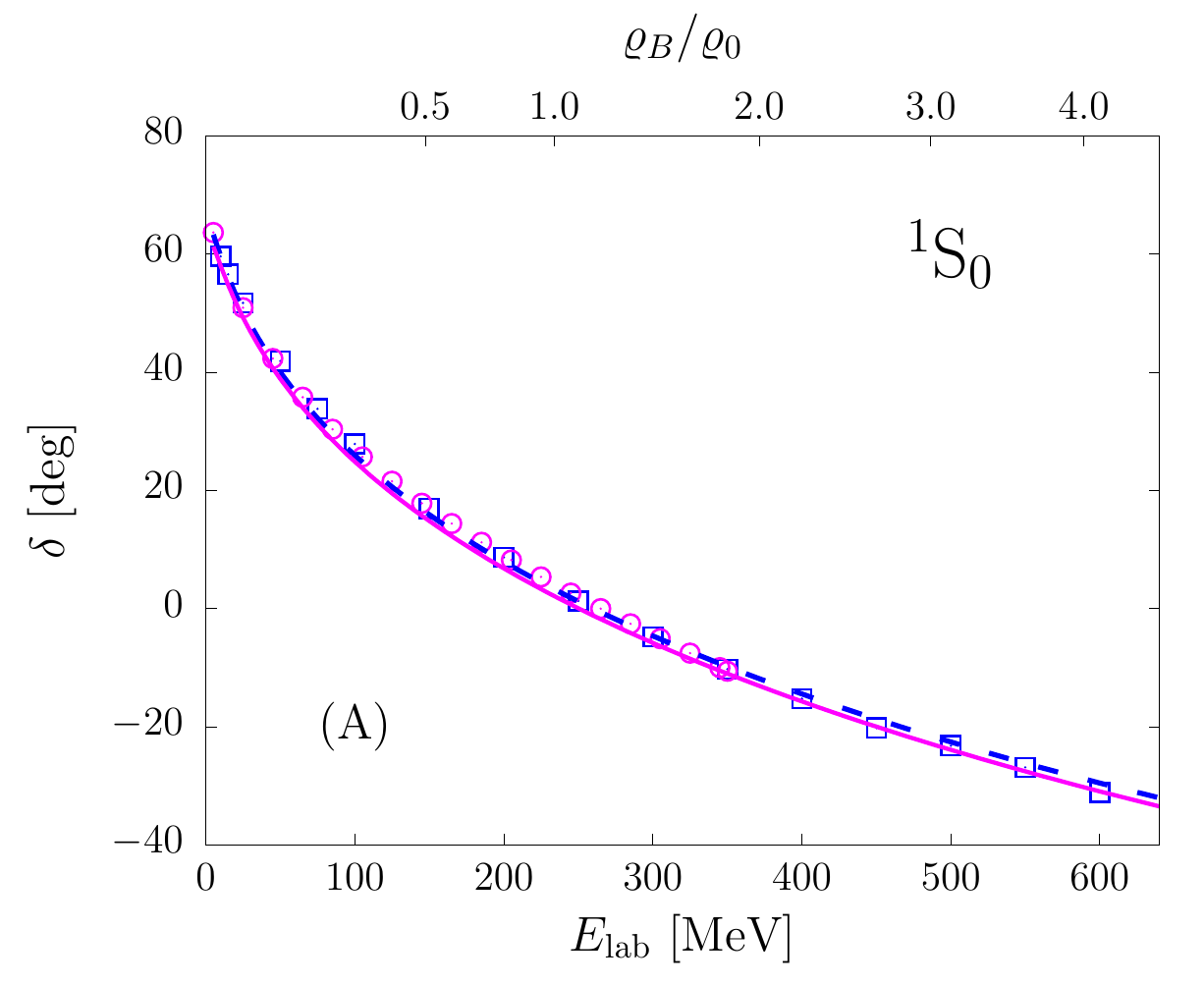}
\includegraphics[scale=0.65]{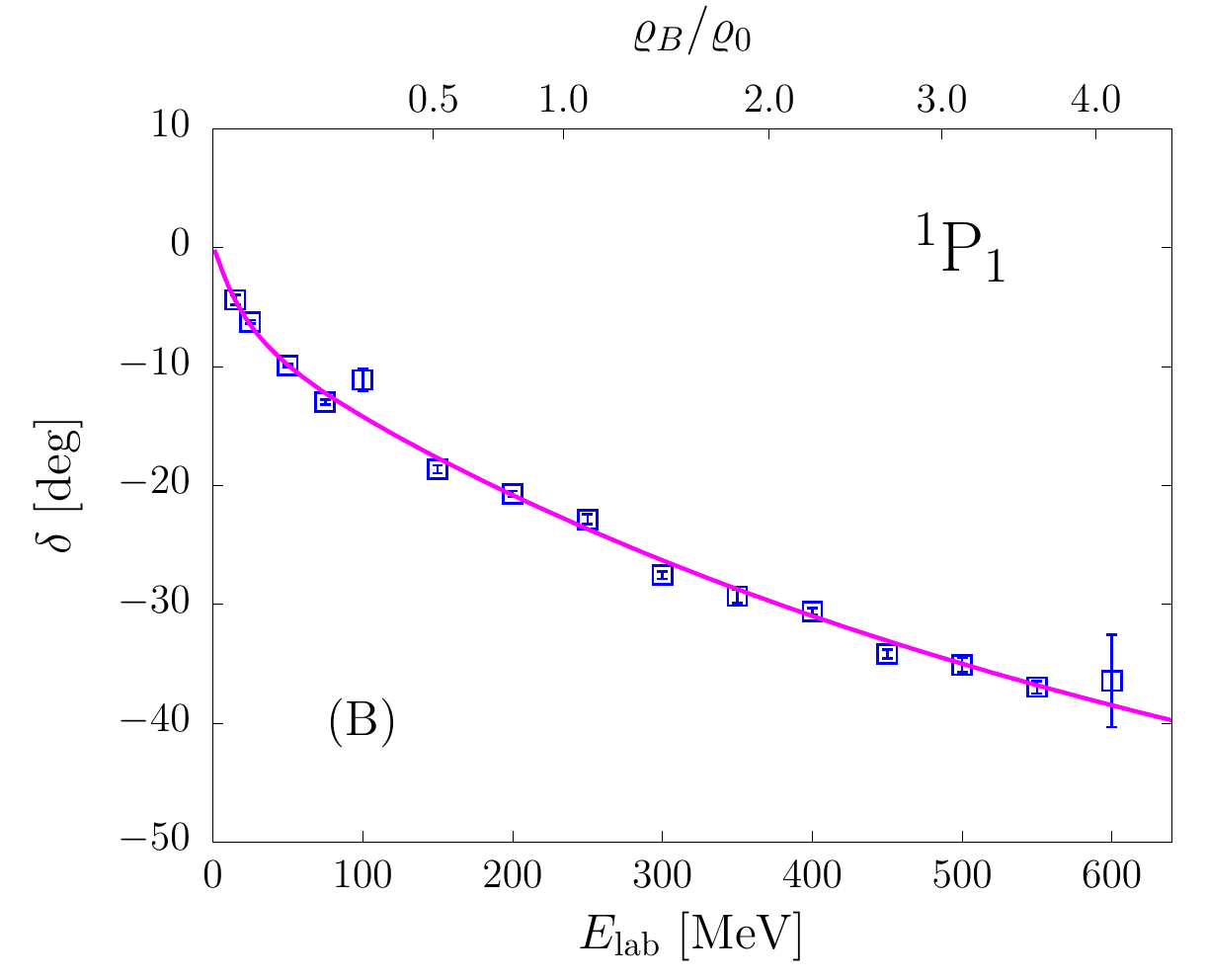} 

\caption{\small Proton-neutron scattering phase shifts in the $^1{\rm S}_0$ (A) and $^1{\rm P}_1$ (B) 
channels, plotted as a function of the kinetic energy of the beam particle in the  
laboratory frame (bottom axes). The corresponding densities of pure neutron matter\textemdash in units of the equilibrium density 
of isospin-symmetric matter, $\varrho_0~=~0.16 \ {\rm fm}^{-3}$\textemdash are given in the top axes. 
The results obtained using the AV6P potential are represented by the solid lines. 
For comparison, the $^1{\rm S}_0$ phase shifts calculated using the full AV18 potential are also shown by the dashed line of 
panel (A).
The data is taken from ~\citet{Nijmegen1}, (circles) \citet{Nijmegen2} (circles), and~\citet{ SAID}(squares).\label{fig:phases}}
\end{figure}

As an example, the energy dependence of the $^1{\rm S}_0$ and $^1{\rm P}_1$ phase shifts is illustrated in Fig.~\ref{fig:phases}. It is apparent that the results obtained using 
the AV6P potential, represented by the solid lines, provide an accurate description of the data reported by ~\citet{Nijmegen1}, \citet{Nijmegen2}, and~\citet{ SAID}
up to beam energies $\sim$~600~MeV, well beyond the pion production threshold, $E_{\rm thr} \approx$~280~MeV.
For comparison, the $^1{\rm S}_0$ phase shifts calculated using the full AV18 potential are also shown by the dashed line of panel (A).

It is very important to realise that the capacity to explain scattering data at large energy is critical to assess the ability 
of a potential model to describe the properties of
nuclear matter in the high-density region, relevant to neutron star physics. To see this, consider a scattering process 
involving two nucleons embedded in the nuclear medium at baryon number density $\varrho_B$. Owing to Pauli's exclusion principle, in the strongly degenerate regime typical of neutron stars collisions predominantly involve particles with energies 
close to the Fermi energy, $E_F$. As a consequence, a relation can easily be established between the kinetic energy of the beam particle in 
the laboratory frame, $E_{\rm lab}$, and the Fermi energy, which in turn is simply related to the density. 
In the case of head-on collisions, the resulting expression reads
\begin{align}
\label{E:rho}
E_{\rm lab} = 4 E_F = \frac{2}{m} \left( \frac{6 \pi^2}{\nu} \varrho_B \right)^{2/3} \ ,
\end{align}
where the degeneracy of the momentum eigenstates, $\nu$, equals to 2 and 4 in pure neutron matter (PNM) and isospin-symmetric matter (SNM), respectively.
The densities of PNM corresponding to beam energies in the range $0\leq~E_{\rm lab}~\leq~630$ MeV, expressed 
in units of the equilibrium density of SNM, $\varrho_0=0.16 \ {\rm fm}^{-3}$,  are given in 
the top axes of Fig.~\ref{fig:phases}. 

Three-body forces are long known to be required to model the interactions of extended composite bodies, such as protons and neutrons, without 
considering their internal structure explicitly~\citep[see, e.g.,][]{Friar}. In nuclear many-body theory, the inclusion of {\it irreducible} three-nucleon (NNN) forces,  described by the potential $V_{ijk}$,  is needed to explain both the observed binding energies of the three-nucleon systems and saturation\textemdash that is, the occurrence of a minimum of the energy per particle at non-vanishing density $\varrho_0$\textemdash of SNM.

The nature of NNN interactions has been first highlighted in the seminal paper of~\citet{Fujita:1957zz}. These authors argued that the most prominent mechanism is the two-pion exchange process, in which one of the nucleons participating in a NN interaction is excited to a $\Delta$ resonance, that then decays in the  aftermath  of the interaction with a third nucleon. Commonly used phenomenological
models of the NNN force, such as the Urbana IX (UIX) potential adopted in this work~\citep{Pudliner:1995wk}, are written in the form
\begin{align}
V_{ijk}=V_{ijk}^{2\pi}+V_{ijk}^{R} \ ,
\end{align}
where $V_{ijk}^{2\pi}$ is the attractive Fujita-Miyazawa term, while $V_{ijk}^{N}$ is a purely phenomenological repulsive term.

Like most phenomenological potentials, the UIX model involves two parameters. The strength of the two-pion exchange 
contribution,  $V_{ijk}^{2\pi}$, is fixed in such a way as to reproduce the observed ground-state energies of 
\isotope[3][]{He} and \isotope[4][]{He}, while that of the isoscalar repulsive contribution, $V_{ijk}^{R}$, is adjusted to obtain the saturation density of SNM inferred from extrapolation of nuclear data.
It is remarkable, however, that saturation of SNM at density $\varrho_B \sim \varrho_0$ is also 
predicted by a potential tuned only to reproduce the properties of the few-nucleon systems~\citep{TM,Lovato:3BF}.

Nuclear Hamiltonians constructed combining the AV18 NN potential and a phenomenological NNN potential, such as the 
UIX model, have been shown to possess a remarkable predictive power. The results of Quantum Monte Carlo (QMC) calculations, extensively 
reviewed by~~\citet{QMC}, reproduce the measured energies of the ground and low-lying excited states of nuclei 
with mass number  $A\leq 12$ to few percent accuracy.  The results reported in the present work have been obtained using the 
AV6P+UIX  Hamiltonian. 

Over the past two decades, a great deal of effort has been devoted to the derivation of nuclear potentials within the framework of
$\chi$EFT~\citep[see, e.g., ][and references therein]{Epelbaum,Machleidt}. This formalism, originally proposed 
by~\citet{weinberg1}, is based on the use of effective Lagrangians involving pions and low-momentum nucleons, 
constrained by the broken chiral symmetry of strong interactions. 

The approach based on $\chi$EFT  provides an elegant and systematic scheme, in which the 
nuclear interaction is expanded in powers of a small parameter, e.g. the ratio between the pion mass or the typical nucleon momentum and the scale of chiral symmetry breaking, $\Lambda_\chi~\sim$~0.8 \textendash  \ 1~GeV. In addition, it allows to obtain two-, three-, and many-nucleon potentials in a fully consistent fashion.
Local coordinate-space representations of potentials derived from $\chi$EFT, suitable to carry out nuclear matter calculations using advanced
many-body techniques, have been first derived by~\citet{gezerlis:2013,gezerlis:2014}. More general coordinate-space potential models, in which the appearance of $\Delta$ resonances in intermediate states is explicitly taken into account, have been developed by~\citet{piarulli}.

In principle, nuclear Hamiltonians obtained from $\chi$EFT may be used to carry out calculations of nuclear matter properties relevant to neutron stars. 
However, it has to be kept in mind that, being based on a low-momentum expansion, $\chi$EFT is inherently limited to the description of matter at densities
$\lesssim 2 \varrho_0$.
This problem clearly emerges from the phase-shift analysis of~\citet{gezerlis:2013}, 
showing that their next-to-next-to-leading order (N$^2$LO) potential only describes the data at $E_{\rm lab} \lesssim 150$~MeV.

The potentials of~\citet{piarulli} have been obtained by fitting NN scattering data up to $E_{\rm lab} = 125$ or $200$ MeV, and splitting the strong interaction
component into short- and long-range contributions involving different coordinate-space cutoffs, $R_L$ and $R_T$. The results of this analysis show that a
suitable combination of $R_L$ and $R_T$ allows to obtain a description of scattering data comparable to that provided by the AV18 potential. It should be pointed out, however, that this procedure, while being fully justified on phenomenological grounds, involves a  departure from the original formulation of the approach based on $\chi$EFT. 

\subsection{Renormalisation in coordinate space: the CBF effective interaction} 
\label{sec:veff}

The observation that the central density of atomic nuclei obtained from elastic electron-nucleus scattering 
data~\citep{frois:densities} becomes largely $A$-independent for $A\gtrsim 12$,
its value being $\sim \varrho_0$,
indicates  that NN forces are strongly repulsive at short distance.  This feature has been  qualitatively confirmed by the results of
pioneering lattice calculations based on the fundamental theory of strong interactions: Quantum Chromo-Dynamics, or QCD~\citep{lattice1}.

Owing to the presence of the repulsive core, 
the matrix elements of the NN potential between eigenstates of the non interacting system are large, 
and standard many-body perturbation theory cannot be used to carry out calculations of nuclear properties.

A time-honored theoretical approach to overcome the above problem\textemdash known as $G$-matrix perturbation theory\textemdash is based on the replacement of the bare NN potential with a well-behaved operator describing NN scattering 
in the nuclear medium. The lowest order approximation of the resulting expansion
has been extensively employed in early  studies of cold nuclear matter~\citep[see, e.g.,][]{day_rmp,Bethe}. More recent developments, allowing the inclusion of higher order terms and the treatment of matter at  finite temperature, are thoroughly reviewed in the volume edited by M.~\citet{Baldo}. 
 
An alternative scheme to determine effective interactions suitable to carry out perturbative calculations\textemdash founded  
on the CBF formalism and the cluster expansion technique~\citep[see][and references therein]{CLARK197989,FF:CBF}\textemdash has been  
proposed  in the early 2000s by~\citet{cowell:2003,cowell:2004}, and further developed by~\citet{bv}, \citet{LLB,LBGL}, and~\citet{eos0}.
    
The CBF effective interaction, that can be written in the form 
\begin{equation}
\label{def:veffp}
v_{ij}^\text{eff}=\sum_{p=1}^6 v^{\text{eff}\,, p}(r_{ij}) O^{p}_{ij} \ ,
\end{equation}
with the $O^{p}_{ij}$ given by Eqs.\eqref{av18:2} and \eqref{S12},  is {\em defined} by the equation
\begin{equation}
\label{def:veff}
\langle H \rangle = \langle\Psi_0 | H | \Psi_0 \rangle = T_F + \langle \Phi_0 | \sum_{i<j} v_{ij}^\text{eff} | \Phi_0\rangle \ .
\end{equation}
Here, $| \Phi_0\rangle$ and $T_F$ denote the ground state of the non interacting Fermi gas at density $\varrho_B$ and the 
corresponding energy, respectively, while $H$ is the nuclear Hamiltonian of Eq.~\eqref{H:A}.
The {\it correlated} ground state, $| \Psi_0\rangle$, is obtained from the corresponding Fermi gas state $| \Phi_0\rangle$ through the transformation
\begin{equation}
\label{def:corrfun}
|\Psi_0 \rangle \equiv \frac{{F}|\Phi_0\rangle}{\langle \Phi_0 | {F}^\dagger {F} |\Phi_0\rangle^{{1/2}}} \ ,
\end{equation}
where the operator ${F}$ 
is a product of two-body correlation operators, whose structure is chosen in such a way as to  reflect the complexity of NN interactions. The resulting expression, 
to be compared to Eq.~\eqref{eq:NN_1}, reads
\begin{equation}
\label{corr1}
{F}  \equiv \mathcal{S} \prod_{i<j} F_{ij}   \ ,
\end{equation}
with
\begin{equation}
\label{corr11}
F_{ij}=\sum_{p=1}^6 f^p(r_{ij}) O^{p}_{ij}  \ ,
\end{equation}
where the $f^p$ are NN correlation functions, to be discussed below.
Note that the inclusion of the operator ${\mathcal S}$, appearing in the right-hand side of Eq.~\eqref{corr1}, is needed to symmetrise the product 
under particle exchange, because, in general, $[O^{p}_{ij},O^{p}_{jk}] \neq 0$.

The determination of the effective interaction from Eq.~\eqref{def:veff} is based on the cluster expansion of the left-hand side, leading to 
\begin{align}
\label{cluster}
\langle H \rangle = T_F + \sum_n (\Delta E)_n = T_F + \langle \Phi_0 | \sum_{i<j} v_{ij}^\text{eff} | \Phi_0\rangle \ , 
\end{align} 
where $(\Delta E)_n$ denotes the contribution to the Hamiltonian expectation value arising from $n$-nucleon clusters. In order to explicitly take into account three-nucleon forces, the effective interaction employed in this work has been derived including terms with $n=$ 2 and 3.

\begin{figure}[ht!]
\centerline{ \includegraphics[scale=0.65]{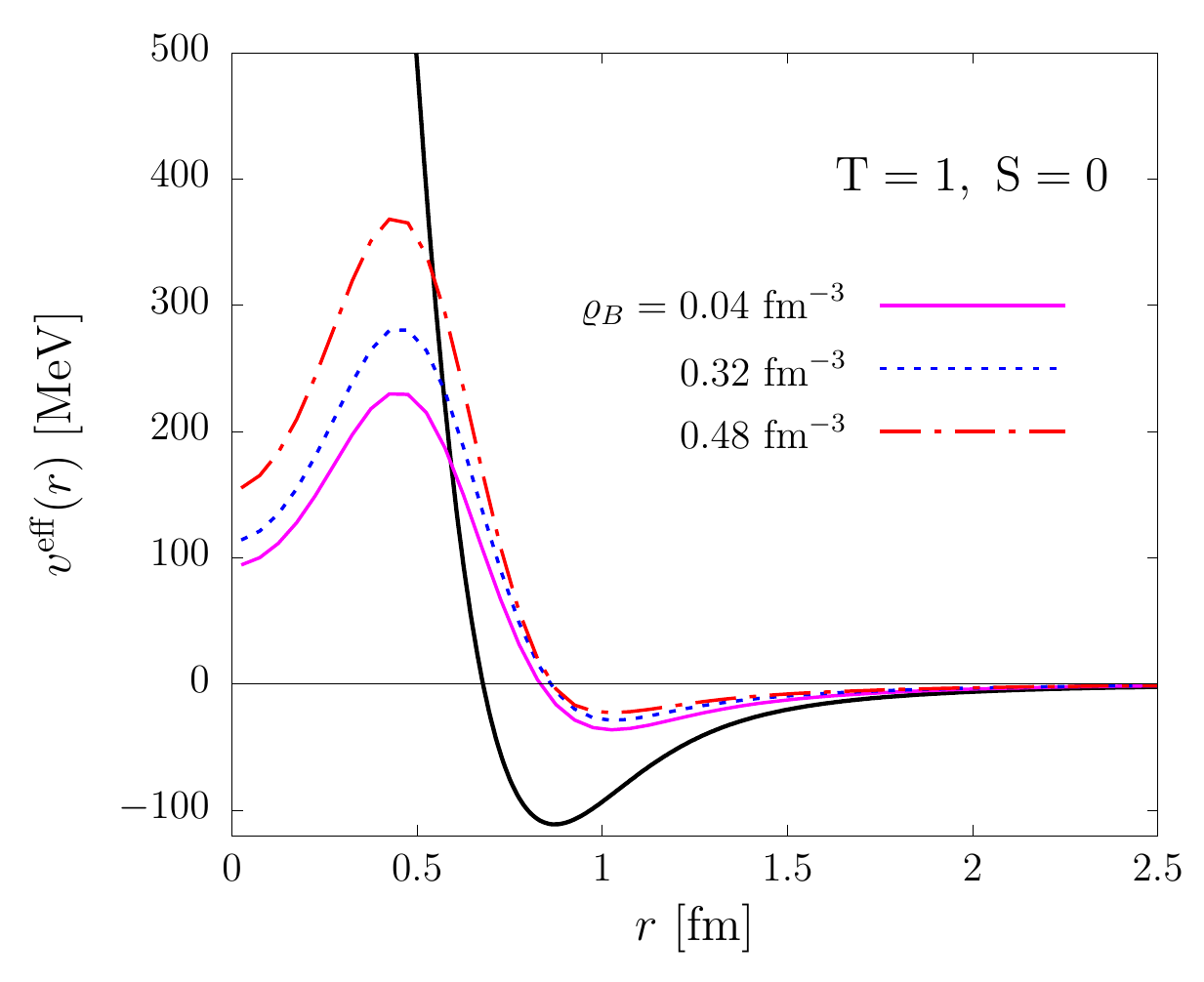} }
\caption{Radial dependence of the CBF effective potential  in the  $S=0$, $T=1$  channel.
The solid, dashed, and dot-dash lines correspond to baryon density $\varrho_B =$ 0.04, 0.32 and 0.48 ${\rm fm}^{-3}$. 
For comparison, the thick solid line shows the bare AV6P potential. \label{densdep}}
\end{figure}

The radial  dependence of the correlation functions is obtained solving a set of Euler-Lagrange equations derived from functional 
minimisation of the two-body cluster approximation to $\langle H \rangle$~\citep{bob_vijay_rmp}. The range of the $f^p(r_{ij})$ is fixed in such a way as to simultaneously reproduce the ground-state energies of PNM and SNM obtained from highly accurate many-body calculations, carried out using the 
Auxiliary-Field-Diffusion-Monte-Carlo (AFDMC) technique or the variational approach referred to as Fermi-Hyper-Netted-Chain/Single-Operator-Chain (FHNC/SOC)~\citep{eos0}.

Note that the CBF effective interaction depends on density through  both dynamical correlations, described by the operator $F_{ij}$, and statistical correlations, arising from the antisymmetric nature  
 of the state $ | \Phi_0\rangle$. The radial dependence of $v^{\rm eff}$ in the $S=0$ and $T=1$ channel at baryon density $\varrho_B =$ 0.04, 0.32 and 0.48 fm$^{-3}$ is displayed in Fig.~\ref{densdep}.

\begin{figure}[ht!]
\centerline{ \includegraphics[scale=0.50]{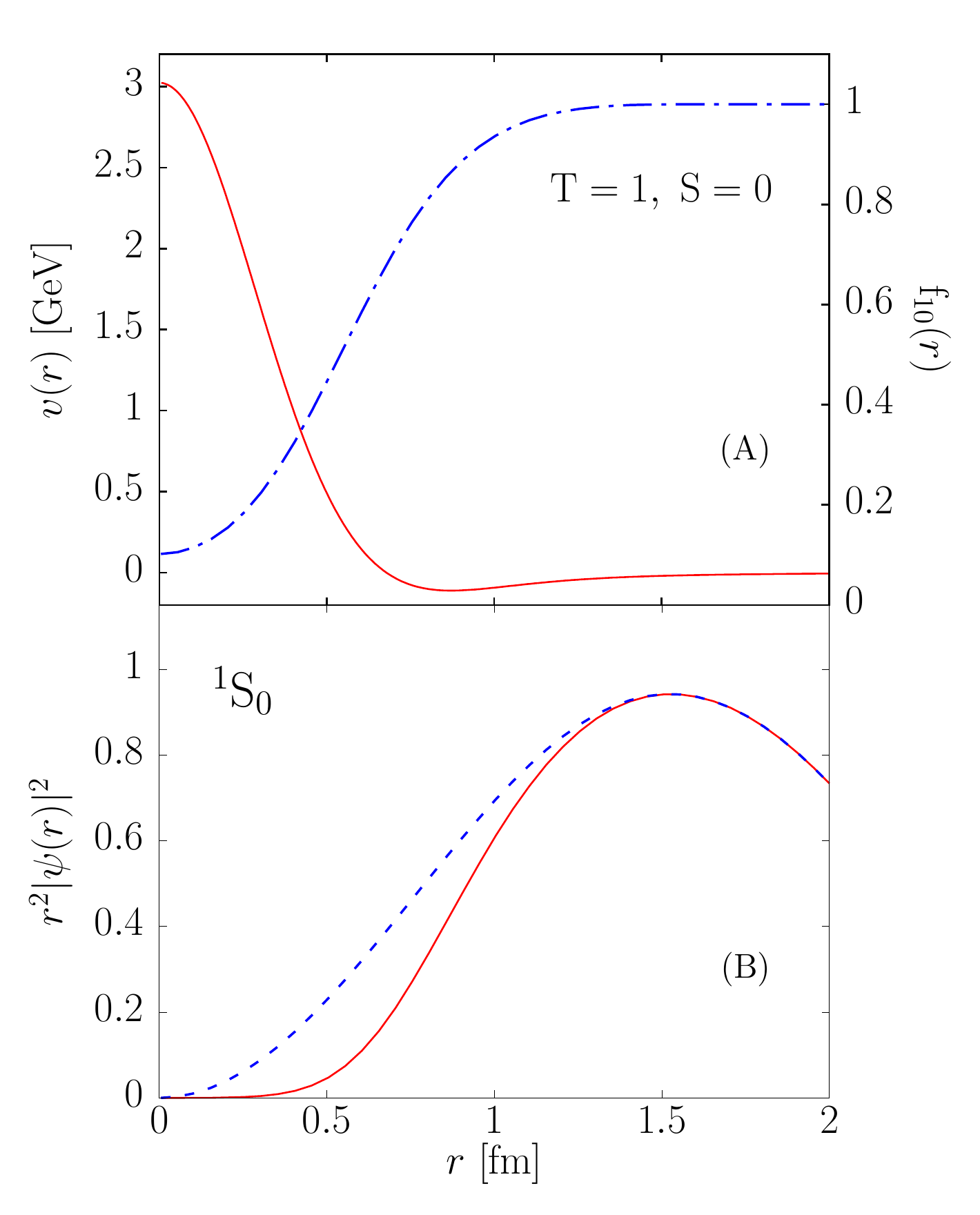} }
\caption{(A): radial dependence of the AV6P potential in the   $S=0$, $T=1$ channel (solid line, left axis) and of the corresponding correlation function (dot-dash line, right axis).
 (B): radial probability associated with the wave function of a correlated  pairs in $^1$S$_0$ state embedded in nuclear matter at equilibrium density (solid line). 
 The dashed line represents the same quantity in the absence of correlations. The probabilities are given in arbitrary units, and the momentum 
 appearing in Eq.~\eqref{corr:wf} is set to $k = \sqrt{3/5} \ k_F$. \label{screen}}
\end{figure}

The short-distance behaviour of the $f^p$\textemdash shaped by the strongly repulsive core of the NN potential\textemdash brings about a strong suppression of 
the probability of finding two nucleons at relative distance $r \lesssim~1$~fm. To see this, consider a nucleon pair in the state
of total spin and isospin $S=0$ and $T=1$, angular momentum $\ell=0$ and momentum {\bf k}, embedded in in nuclear matter at equilibrium density. 
Its relative motion is described by the wave function 
\begin{align}
\label{corr:wf}
\psi(r) = f_{10}(r) \frac{ \sin{kr} }{kr} \  ,
\end{align}
where the spin-isospin projected correlation function $f_{10}$ is a linear combination of the $f^{p}$ of Eq.~\eqref{corr11} with $p \leq 4$. 
The radial dependence of the NN potential and the correlation function $f_{10}$ is displayed in panel (A) of 
of Fig.~\ref{screen}, while panel (B) shows a comparison between 
the probability associated with the wave function of Eq.~\eqref{corr:wf}, $r^2 | \psi(r)|^2$, and the corresponding quantity in the absence of correlations. 


A great deal of effort has been also devoted to the derivation of soft effective interactions\textemdash suitable to carry out
perturbative calculations of nuclear matter in the Fermi gas basis\textemdash  
from renormalisation-group (RG) evolution of potentials obtained within $\chi$EFT~\citep[see, e.g.,][]{RG, RG2}.
It has to be emphasised that the RG approach is conceptually equivalent to the one based on the CBF formalism. Within the RG scheme, 
screening of the repulsive core of the NN interaction is obtained by integrating out high-momentum components, and the evolution  
is driven by the value of the momentum cutoff. Using the CBF formalism, on the other hand, the same effect is realised in coordinate space 
through the action of the correlation functions, as illustrated in Fig.~\ref{screen}, and evolution is driven by nuclear matter density, determining 
the average distance between matter constituents. However, it has to be kept in mind that\textemdash in view of the limitations discussed in Section~\ref{sec:ham}\textemdash the RG approach, being based on potentials derived from $\chi$EFT, 
is expected to be applicable in a narrow density region.

\section{Finite-Temperature Perturbation Theory} 
\label{finite:T}

The basic assumption underlying our treatment of nuclear matter at $T\neq0$ is that  at low-to-moderate 
temperatures\textemdash typically $T \ll m_\pi$, $m_\pi \approx 140$ MeV being the pion mass\textemdash nuclear 
dynamics, described by the potentials appearing in the Hamiltonian of Eq.~\eqref{H:A},  is largely unaffected by thermal effects. 
In principle, the CBF effective interaction involves an additional temperature dependence associated with the 
correlation functions, since the Fermi distribution appears in the Euler-Lagrange equations determining their shape. However, the results of detailed numerical calculations have shown that thermal modifications 
of the $f^p$ of Eq.~\eqref{corr11} turn out to be negligibly small
up to $T \sim 50$~MeV~\citep{valli:thesis}.
The results reported in this article have been obtained using the zero-temperature effective interaction, involving 
correlation functions computed at $T=0$.

\subsection{Thermodynamics}
\label{thermodynamics} 

All thermodynamic functions of a system in thermal equilibrium at temperature $T$ can be derived from the grand canonical potential, defined as 
\begin{align}
\label{def:Omega}
\Omega = - \frac{1}{\beta} \ln Z \ , 
\end{align}
where $\beta = 1/T$. In the case of a one-component system, the partition function, $Z$, is given by 
\begin{align}
\label{def:Z}
Z = {\rm Tr} \  \Phi \ ,  
\end{align}
with 
\begin{align}
\label{def:Phi}
\Phi = e^{-\beta(H - \mu N)} \ .
\end{align}
In the above equation, $\mu$ is the chemical potential, while  $H$ and 
$N$ denote the Hamiltonian and the particle number operator, respectively\footnote{For simplicity, here we discuss the case of a 
one-component system, such as unpolarised PNM. In a multi-component system, such as
$\beta$-stable matter, Eq.~\eqref{def:Phi} takes the form $\Phi = \exp [ - {\beta ( H - \sum_\lambda \mu_\lambda N_\lambda) ]}$, where the sum is extended to all  particle species.}. 

The basis for the derivation of finite-temperature perturbation theory is provided by the Bloch equation~\cite[see, e.g.,][]{Thouless}  
\begin{align}
\label{bloch}
- \frac { \partial \Phi }{ \partial \beta } = (H - \mu N)\Phi \ , 
\end{align}
to be solved with the boundary condition $\Phi(0) = 1$.

The perturbative expansion of the grand canonical  partition function is easily obtained 
exploiting the formal similarity between Eq.~\eqref{bloch} and the time-dependent Schr\"odinger equation of quantum mechanics, 
and rewriting the Hamiltonian in the form
\begin{align}
\label{def:H}
H & = H_0 + H_I . 
\end{align}

Substitution of Eq.~\eqref{def:H} into the right-hand side of the Bloch equation, leading to 
\begin{align}
\label{def:mH}
\nonumber
- \frac { \partial \Phi }{ \partial \beta } & = [ (H_0 - \mu N) + H_I ]\Phi \\
                                                          & = (H_0^\prime + H_I ) \Phi  \ ,
\end{align}
shows that the formalism of time-dependent perturbation theory can be readily generalised by replacing 
$t \to -i \beta$, and using the operator $H_0^\prime$ to define the appropriate
interaction picture.  

The fundamental relation~\citep[see, e.g.,][]{landau} 
\begin{align}
\label{fundamental}
\Omega = -PV = F - \mu N = E - TS - \mu N  \ ,
\end{align}
provides a link between the grand canonical potential, the pressure $P$, and the free energy $F = E - TS$, with $E$ and $S$ being the energy and  entropy
of the system, respectively. From Eq.~\eqref{fundamental} 
if follows that 
\begin{align}
P = - \frac{\Omega}{V} \ \ \ , \ \ \ S = - \frac{\partial \Omega}{\partial T} \ \ \ , \ \ \  N = - \frac{\partial \Omega}{\partial \mu} \ . 
\end{align}

In the following, we will discuss the application of the above results to a system described by the Hamiltonian
\eqref{def:H}, with 
\begin{align}
\label{def:H0}
H_0 = \sum_k e_k a^\dagger_k a_k \  , 
\end{align}
where, in general 
\begin{align}
\label{def:ek}
e_k = \frac{ {\bf k}^2}{2m} + U_k = t_k + U_k \ ,  
\end{align}
and
\begin{align}
\label{def:HI}
H_I = \sum_{ k, k^\prime, q, q^\prime }  \langle k^\prime q^\prime | v |  k q \rangle
 a^\dagger_{k^\prime} a^\dagger_{q^\prime} a_{q} a_{k}
- \sum_k U_k a^\dagger_k a_k \  .
\end{align}
Here, the label $k$ specifies both the particle momentum and the discrete quantum numbers corresponding to one-particle states, 
$a^\dagger_{k}$ and $a_{k}$ denote creation and annihilation operators, respectively, and $v$ is the
potential describing interparticle  interactions. The single-particle potential $U_k$, which in principle does not 
affect the results of calculations of physical quantities, is  chosen in such a way as to improve the convergence of
the perturbative expansion, or to fulfill specific conditions~\citep[see, e.g.,][]{Baldo} . 

Note that, through the use of the CBF effective interaction discussed in Section~\ref{sec:veff}, the formalism based on the Hamiltonian
defined by Eqs.~\eqref{def:H}~and~\eqref{def:H0}-\eqref{def:HI} allows to take into account two- and three-nucleon interactions
in a consistent fashion.

It has to be pointed out that, according to Eq.~\eqref{fundamental}, the pressure can be written in the form
\begin{align}
\label{HV}
P = \varrho_B \Big( \mu - \frac{F}{N} \Big) \ , 
\end{align}
with $\varrho_B = N/V$, implying that at equilibrium, that is, for $P=0$,  $\mu = F/N$. 
This result can be seen as the generalisation of the
Hugenholtz-Van Hove theorem~\citep{HVH} to the case of non vanishing temperature.

\subsection{Perturbation Theory}
\label{sec:energy}

At first order in $H_I$, the grand canonical potential is given by~\cite[see, e.g.,][]{lejeune:NPA}
\begin{align}
\label{Omega:pert}
\Omega = \Omega _0 + \Omega_1 \ , 
\end{align}
with
\begin{align}
\label{Omega0}
\Omega_0 & = - \frac{1}{\beta} \sum_k \ln \big\{ 1 + e^{-[\beta(e_k - \mu)]} \big\} \ , \\
\label{Omega1}
\Omega_1 & = \frac{1}{2} \sum_{k k^\prime}  \langle k k^\prime | v | k k^\prime \rangle_A \ n_k n_{k^\prime}   
 - \sum_k U_k n_k \ ,
\end{align}
where $ | k k^\prime \rangle_A = | k k^\prime \rangle - | k^\prime k \rangle$ denotes an antisymmetrised two-particle state, 
and $n_k$ is the Fermi distribution, given by
\begin{align}
n_k = \big[ 1 + e^{\beta(e_k - \mu)} \big]^{-1} \ . 
\label{fermidist:1}
\end{align}
From Eqs.\eqref{Omega0} and \eqref{Omega1} it follows that the free energy per particle 
\begin{align}
\label{F1}
\frac{F}{N} = \frac{1}{N} ( \Omega_0 + \Omega_1) + \mu   \ , 
\end{align}
can be cast in the form
\begin{align}
\nonumber
\label{F2}
\frac{F}{N}  & = \frac{1}{N} \Big\{ \sum_k  t_k n_k + \frac{1}{2} \sum_{k, k^\prime} 
 \langle k k^\prime | v | k k^\prime \rangle_A \ n_k n_{k^\prime} 
 + \frac{1}{\beta} \sum_k \big[ n_k \ln n_k + (1-n_k) \ln (1-n_k) \big] 
 \nonumber
+ \mu \Big(1 - \frac{1}{N} \sum_k n_k \Big) \Big\} \ .
\end{align}
In principle, for any assigned values of temperature  and chemical potential, the above equations provide a scheme for the determination of the equation of 
state of nuclear matter at finite temperature,  $P = P(\mu,T)$. In view of the fact that baryon number is conserved by all known interactions, however, in nuclear matter it is convenient to use baryon density as an independent variable, and determine the chemical 
potential from the relation
\begin{align}
\varrho_B = - \frac{1}{V} \frac{\partial}{\partial \mu} \big( \Omega _0 + \Omega_1 \big) \ .
\end{align}

In the $T \to 0$ limit
\begin{align}
n_k \to \theta(\mu - e_k) \ \ \ , \ \ \ \frac{\partial n_k}{\partial \mu} \to \delta(e_k - \mu)  \ , 
\end{align} 
and the chemical potential is given by  $\mu = e_{k_F}$, with 
the Fermi momentum being defined as $k_F = \big( 6 \pi^2 \varrho_B / \nu \big)^{1/3}$.

For $T\neq0$ and density-dependent potentials, thermodynamic consistency is not trivially achieved in 
perturbative calculations. A clear manifestation of this difficulty is the mismatch between the value of pressure 
obtained from Eq.~\eqref{HV} and the one resulting from the alternative\textemdash although in principle equivalent\textemdash thermodynamic expression
\begin{align}
P= - \frac{\partial F}{\partial V} = \varrho_B^2 \frac{\partial}{\partial \varrho_B} \frac{F}{N} \ .   
\end{align}

A procedure fulfilling the requirement of thermodynamic consistency by construction can be derived from 
a variational approach, based on minimisation of the trial grand canonical potential~\citep{heyer:PLB}
\begin{align}
\widetilde{\Omega} =  \sum_k  t_k n_k & + \frac{1}{2} \sum_{k, k^\prime} 
 \langle k k^\prime | v | k k^\prime \rangle_A \ n_k n_{k^\prime} \\
 \nonumber
 & + \frac{1}{\beta} \sum_k \big[ n_k \ln n_k + (1-n_k) \ln (1-n_k) \big] , 
\end{align}
with respect to the form of $n_k$.  Note that the above 
expression\textemdash the use of which is fully legitimate in the variational context\textemdash can also be obtained in first order perturbation theory neglecting terms involving $\partial \Omega_1 / \partial T$ and 
$\partial \Omega_1 / \partial \mu$~\citep{lejeune:NPA}.

The condition 
\begin{align}
\frac{\delta \widetilde{\Omega}  }{\delta n_k} = 0  \ , 
\end{align}
turns out to be satisfied by the distribution function
\begin{align}
\label{Fermi:consistent}
n_k = \big\{ 1 + e^{\beta[(t_k + U_k + \delta e) - \mu]} \big\}^{-1} \ , 
\end{align}
with
\begin{align}
\label{HF}
U_k = \sum_{k^\prime} 
 \langle k k^\prime | v | k k^\prime \rangle_A \ n_{k^\prime} \ , 
\end{align}
and
\begin{align}
\label{def:deltae}
\delta e= \frac{1}{2} \sum_{k, k^\prime} \langle k k^\prime | \frac{ \partial v}{\partial \varrho_B}  | k k^\prime \rangle_A \ n_k n_{k^\prime} \ .
\end{align}

Within this scheme, that reduces to the standard Hartee-Fock approximation in the case of density-independent potentials, all thermodynamic functions 
at given temperature and baryon density can be consistently obtained using the distribution $n_k$ of Eq.~\eqref{Fermi:consistent}. Note, however, that, because both $U_k$ and $\delta e$
depend on $n_k$, see Eqs.~\eqref{HF} and \eqref{def:deltae}, calculations must be carried out self-consistently, applying an iterative procedure. 

\section{Equation of State of Hot Nuclear Matter}
\label{EOS}

We will consider nuclear matter at fixed baryon density
\begin{equation}
\varrho_B=\sum_\lambda \varrho_\lambda= \sum_\lambda x_\lambda \varrho_B \ ,
\label{eq:rho_asym}
\end{equation}
where the index $\lambda=1,2,3,4$ labels spin-up protons, spin-down protons, spin-up neutrons and spin-down neutrons, respectively, 
and  the corresponding densities are denoted $\varrho_\lambda = x_\lambda \varrho_B$. 
Even though our formalism is completely general, in the following  we will restrict ourselves to 
the case of unpolarised matter, in which $x_1=x_2$ and $x_3=x_4$.

\subsection{Free energy per nucleon and single particle spectrum}
\label{sec:mstar}

Following the procedure described in Sections~\ref{thermodynamics} and \ref{sec:energy}, the energy per nucleon at temperature $T$
can be obtained from 

\begin{align}
\label{def:E}
\frac{E}{N}(\rho_B,T)  = \sum_{ \lambda} x_\lambda \sum_{ {\bf k} }\frac{ {\bf k}^2 }{2m} \ n_\lambda(k,T) + \frac{ \varrho_B }{2} \sum_{\lambda \lambda^\prime} x_\lambda x_{\lambda^\prime} 
\int d^3x \left[ V^{D}_{\lambda \lambda^\prime}({\bf x})  - V^{E} _{\lambda \lambda^\prime}({\bf x}) L_\lambda(x,T) L_{\lambda^\prime}(x,T)  
 \right] \  .
\end{align}
In the above equation  
\begin{align}
V^{D}_{\lambda \lambda^\prime}({\bf x})  = \sum_{p} v^{{\rm eff},p}(x) \langle \lambda\lambda^\prime | O^{p}_{12} | \lambda \lambda^\prime\rangle \ , \\  
V^{E}_{\lambda \lambda^\prime}({\bf x})  = \sum_{p} v^{{\rm eff},p}(x) \langle \lambda \lambda^\prime | O^{p}_{12} | \lambda^\prime \lambda\rangle \ , 
\end{align}
with the functions $v^{{\rm eff},p}$ and the operators $O^p$ defined in Eqs.~\eqref{eq:NN_1} and \eqref{av18:2}, respectively, whereas 
$| \lambda \lambda^\prime\rangle$ denotes a  two-nucleon state in spin-isospin space. 

The temperature dependence of the interaction contributions is contained in the generalised Slater functions $L_\lambda(x)$, defined as
\begin{align}
L_\lambda(x,T) = \frac{1}{\rho_\lambda} \int \frac{d^3 k}{(2 \pi)^3} e^{ i {\bf k} \cdot {\bf x} } n_\lambda(k,T) \ , 
\end{align}
with the Fermi distribution given by
\begin{align}
n_\lambda(k,T) =   \big\{ 1 + e^{( e_{k,\lambda}  - \mu_\lambda )/T } \big\}^{-1} \ , 
\end{align}
where
\begin{align}
\label{def:spectrum}
e_{k,\lambda}  = \frac{{\bf k}^2}{2m} 
 &+  \sum_{\lambda^\prime} \varrho_{\lambda^\prime} \int d^3x \big[ V^D_{\lambda \lambda^\prime}(x) - V^D_{\lambda \lambda^\prime}(x)
j_0(kx) L_{\lambda^\prime}(x,T) \big] \\
\nonumber
& + \sum_{\lambda \lambda^\prime}  \varrho_{\lambda}  \varrho_{\lambda^\prime} \int d^3x 
\Big\{ \big[ \frac{\partial}{\partial \varrho_B} V^D_{\lambda \lambda^\prime}(x) \big] - 
\big[ \frac{\partial}{\partial \varrho_B} V^E_{\lambda \lambda^\prime}(x) \big] L_\lambda(x,T) L_{\lambda^\prime}(x,T) \Big\} \ , 
\end{align}
and $j_0(z) = \sin z/z$.

In the $T \to 0$ limit, $\mu_\lambda = e_{k_{F_\lambda}}$, with $ k_{F_\lambda} = (6 \pi^2 \rho_\lambda)^{1/3}$,  and 
$L_\lambda(x)$ reduces to the ordinary Slater function $\ell(k_{F_\lambda} x)$, with  $\ell(z)~=~3( \sin z - z \cos z)/z^3$.
At $T \neq 0$, however,  $E/N$ and $e_{k,\lambda}$ must be computed self consistently, and  the chemical potentials $\mu_\lambda$ are
determined by the conditions
\begin{align}
\label{det:chempot}
\int \frac{d^3 k}{(2 \pi)^3} n_\lambda(k,T)  = \varrho_\lambda \ .
\end{align}

The free energy per nucleon
\begin{align}
\label{def:F}
\frac{F}{N}(\rho_B,T) = \frac{1}{N} \left[ E(\rho_B,T) - T S(\rho_B,T) \right] \ , 
\end{align}
is obtained combining the above results with the corresponding expression of the entropy per nucleon
\begin{align}
\label{def:S}
\frac{S}{N}(\rho_B,T) = - \frac{1}{\varrho_B} \sum_\lambda \int \frac{d^3 k}{(2 \pi)^3}  \big\{ n_\lambda(k,T) \ln n_\lambda(k,T)  
 + [1 - n_\lambda(k,T)] \ln [1 - n_\lambda(k,T)] \big\} \ .
\end{align}
\begin{figure*}[ht!]
\includegraphics[scale=0.720]{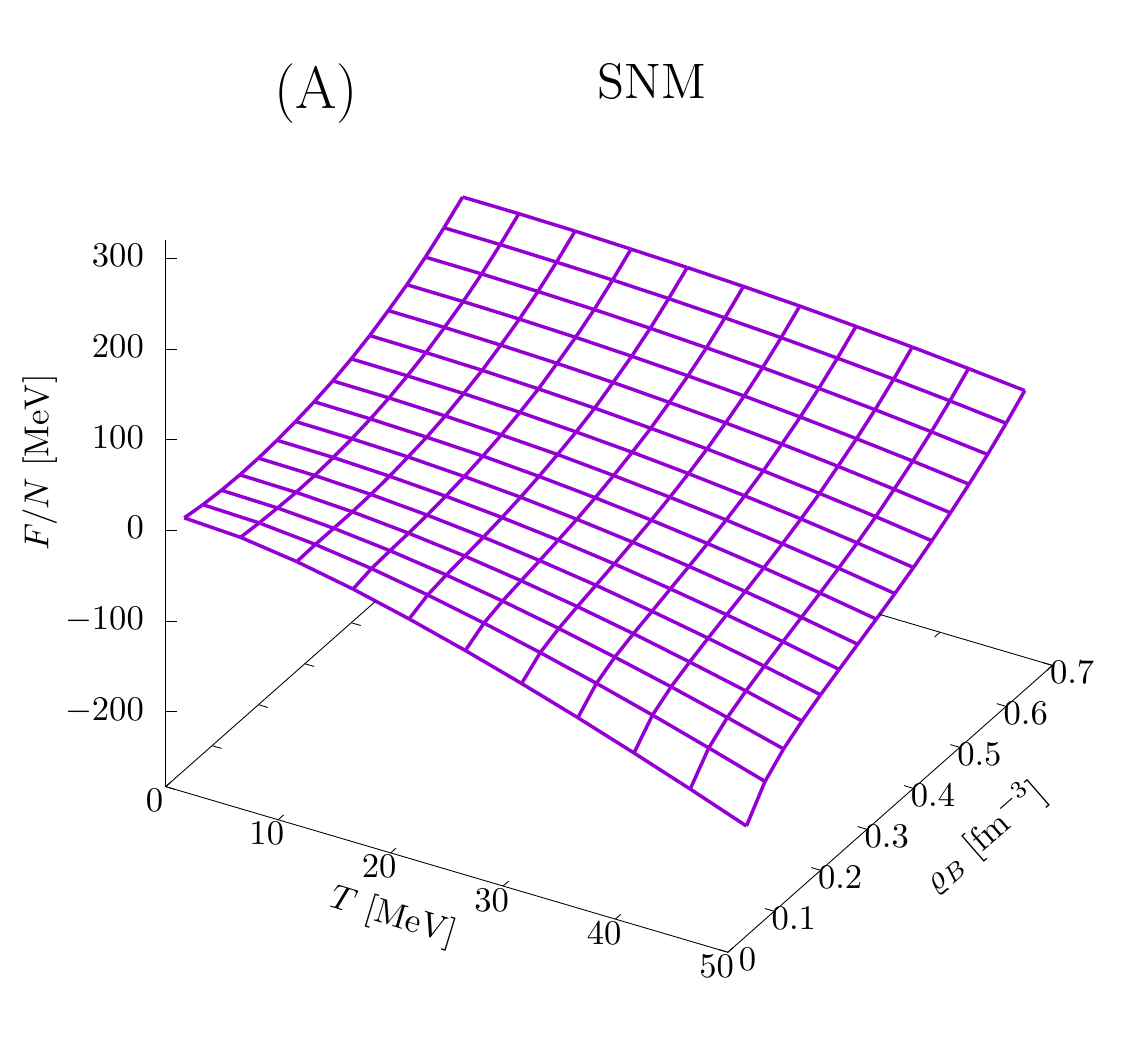} 
\includegraphics[scale=0.745]{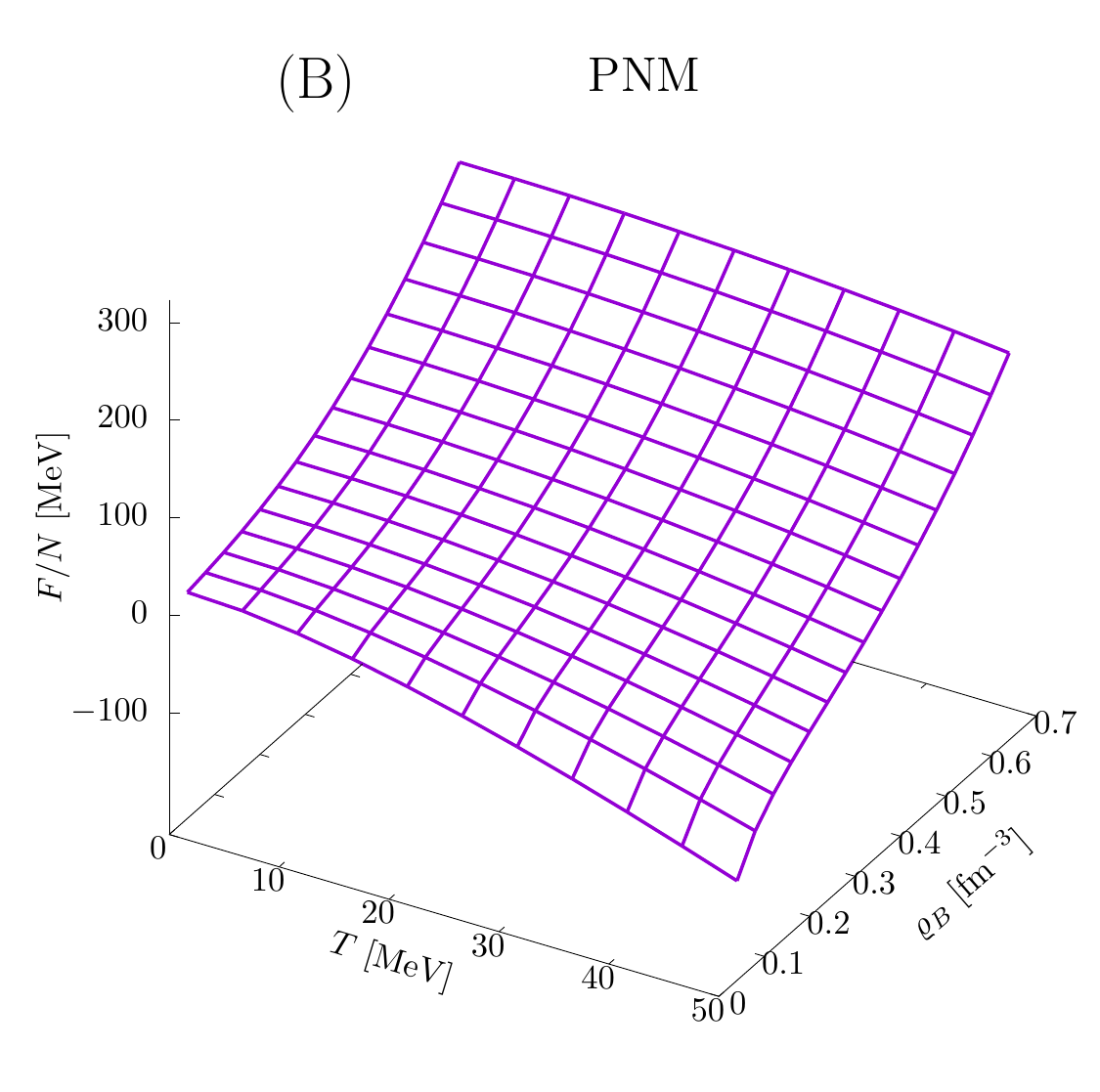}
\caption{Density and temperature dependence of the free energy per nucleon of SNM (A) and PNM (B), computed using Eqs.~\eqref{def:E}, \eqref{def:F}, and \eqref{def:S}, with  the CBF effective interaction described in Section~\ref{sec:veff}.  
 \label{F:3D}}
\end{figure*}

Figure \ref{F:3D} shows the density and temperature dependence of the free energy per 
nucleon of SNM and PNM, corresponding to proton fraction ${\rm Y}_p = x_1 + x_2  = 0.5 $ and 0, respectively, obtained from the  procedure described above using the CBF effective interaction.

\begin{figure}[h!]
\begin{center}
\includegraphics[scale=0.720]{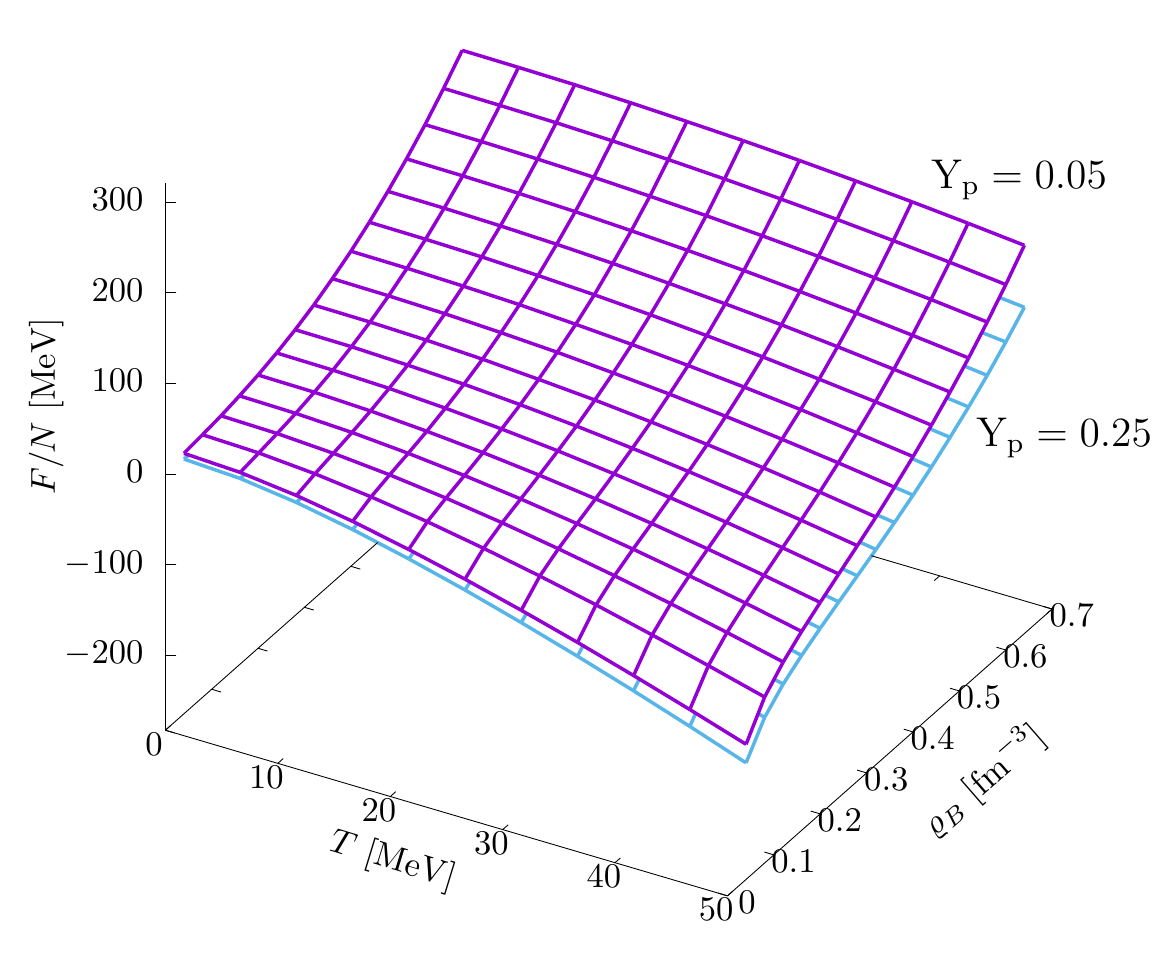}
\end{center}
\caption{Density and temperature dependence of the free energy per nucleon of of isospin-asymmetric matter with 
proton fraction ${\rm Y}_p = 0.05$ and 0.25, computed using Eqs.~\eqref{def:E}, \eqref{def:F}, and \eqref{def:S} with  the CBF effective interaction described in Section~\ref{sec:veff}.  
 \label{xdep}}
\end{figure}

The formalism employed in this work allows to carry out calculations of the properties of 
isospin-asymmetric matter for any values of the proton fraction ${\rm Y}_p$. The smooth ${\rm Y}_p$-dependence of the 
free energy per nucleon is illustrated in Fig.~\ref{xdep}, showing results corresponding to 
${\rm Y}_p$ = 0.05 and 0.25. 

The energy per nucleon of asymmetric matter at zero temperature is often derived from an expansion in powers 
of the neutron excess $\delta = 1 - 2Y_p$. The resulting expression
\begin{align}
\frac{E}{N}(\varrho_B,\delta) \approx  \frac{E}{N}(\varrho_B,0) + E_{\rm sym}(\varrho_B) \delta^2 \ , 
\label{quadratic:0}
\end{align}
with the symmetry energy being given by $E_{\rm sym}(\varrho_B) =  [ E(\varrho_B,1) - E(\varrho_B,0) ]/N$, 
has been shown to provide a remarkably accurate approximation to the results of
calculations in which the dependence on $Y_p$ is taken into account at microscopic level~\citep[see, e.g.,][]{eos0}.  
To test the validity of  the generalisation of Eq.~\eqref{quadratic:0} to finite temperatures\textemdash previously discussed by, e.g.,~\citet{camelio:thesis} and~\citet{Fiorella:symmetry}\textemdash 
we have compared the ${\rm Y}_p$ 
dependence of the free energy per nucleon at different values of $T$ to the predictions of 
the quadratic interpolation formula
\begin{align}
\frac{F}{N}(\varrho_B,T,\delta) =  \frac{F}{N}(\varrho_B,T,0) + F_{\rm sym}(\varrho_B,T) \delta^2 \ , 
\label{quadratic:T}
\end{align}
with
\begin{align}
F_{\rm sym}(\varrho_B,T) = \frac{1}{N} \left[  F(\varrho_B,T,1) - F(\varrho_B,T,0) \right] \ .
\label{Esym:T}
\end{align}

\begin{figure}[ht!]
\begin{center}
\includegraphics[scale=0.650]{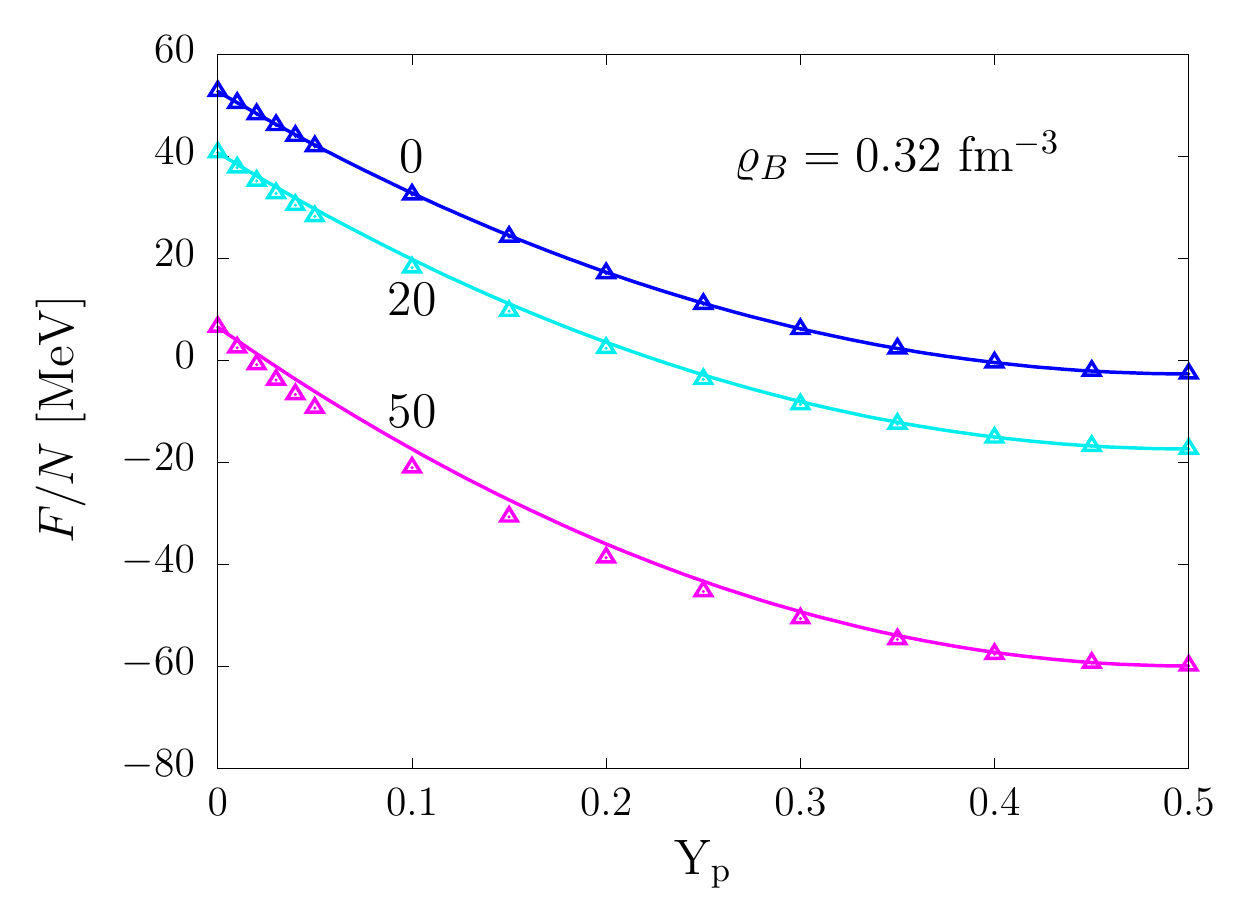}
\end{center}
\caption{Free energy per nucleon of matter at baryon density $\varrho_B = 0.32 \ {\rm fm}^{-3}$, plotted as a function of the proton fraction. The triangles have been obtained using Eqs.~\eqref{def:E}, \eqref{def:F}, and \eqref{def:S} with  the CBF effective interaction described in Section~\ref{sec:veff}, whereas the solid lines show the results of the quadratic approximation discussed in the text. The curves are labelled according to temperature, expressed in units of MeV.   
 \label{quad:xdep}}
\end{figure}

The results shown in Fig.~\ref{quad:xdep}, corresponding to a representative baryon density $\varrho_B = 0.32 \ {\rm fm}^{-1}$, indicate that the 
excellent agreement observed at $T=0$ tends to slowly deteriorate as the temperature increases. However, at $T$ as high as  50 MeV the deviations\textemdash which are vanishing by construction at  $Y_p =$ 0 and 0.5\textemdash turn out to be  still limited to $\sim 20$\% at $Y_p \gtrsim 0.1$.

\subsection{Pressure and thermodynamic consistency}
\label{sec:pressure}

As pointed out in Sec.~\ref{sec:energy}, in the case of density-dependent effective interactions 
thermodynamic consistency, expressed by the condition
\begin{align}
\varrho_B^2 \frac{\partial}{\partial \varrho_B} \frac{F}{N} = 
\varrho_B \left[ \mu_p Y_p + \mu_n (1-Y_p) - \frac{F}{N} \right] \ ,
\label{therm:cons}
\end{align}
where $\mu_n$ and $\mu_p$ are the neutron and proton chemical potentials,
 requires the inclusion of a correction to the Hartree-Fock single-particle spectrum, given by the second line in the
right-hand side of Eq.~\eqref{def:spectrum}.

\begin{figure}[ht!]
\begin{center} 
\includegraphics[scale=0.6750]{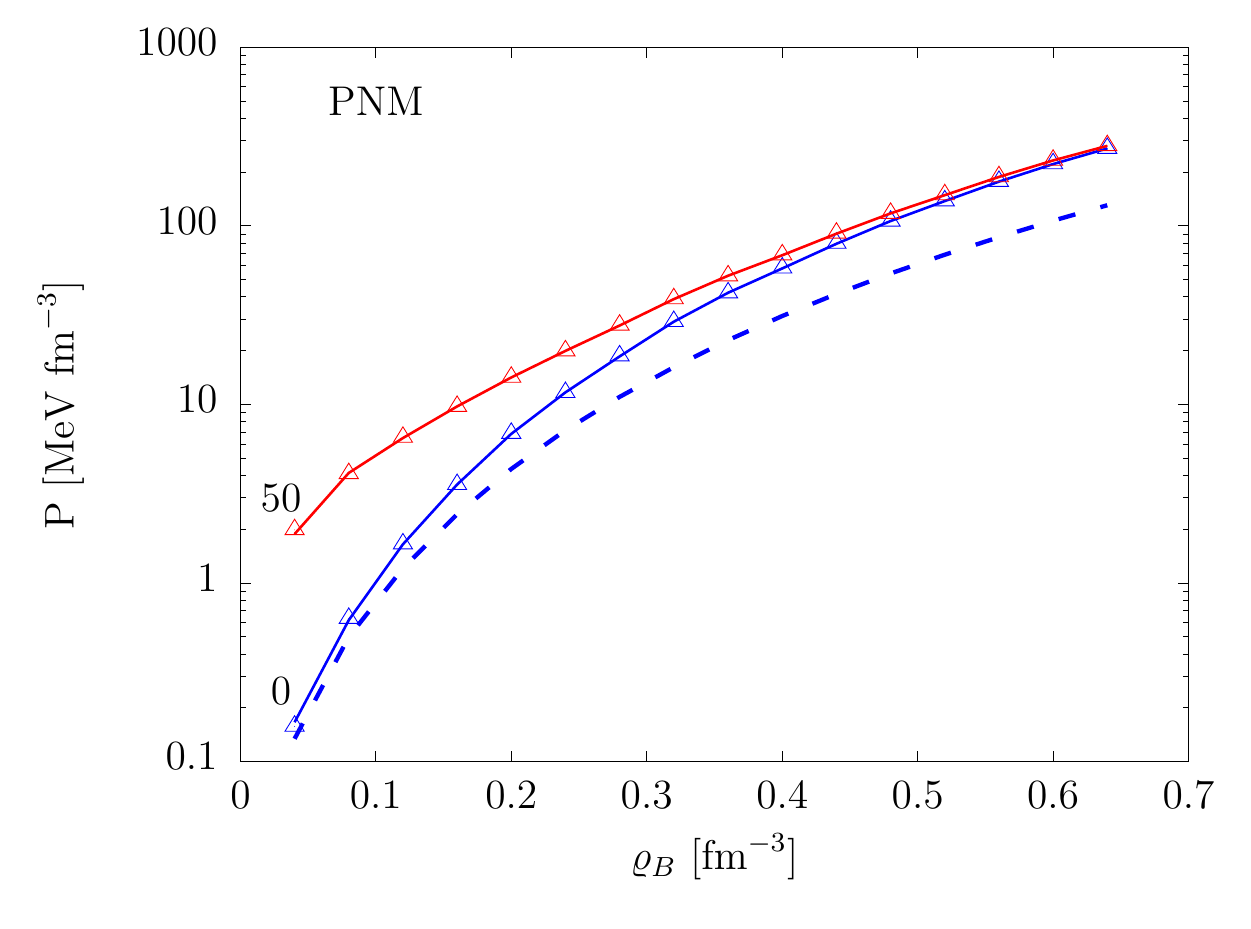}  
\end{center}
\caption{Density dependence of the pressure of PNM at $T=0$ and 50 MeV. Solid lines and triangles correspond to the results 
obtained from the left- and right-hand side of Eq.~\eqref{therm:cons}, respectively. The dashed line represents the results 
obtained neglecting the self-consistency correction to the single nucleon spectrum. 
\label{cons:therm}}
\end{figure}

The thermodynamic consistency of our formalism is illustrated in Fig.~\ref{cons:therm}, showing the pressure of PNM
at $T=$ 0 and 50 MeV. Solid lines and triangles correspond to results 
obtained from the left- and right-hand side of Eq.~\eqref{therm:cons}, respectively. For comparison, the pressure of 
PNM at $T=50$ MeV computed omitting the self-consistency correction to the single-particle spectrum is
shown by the dashed line. It should also be pointed out that the effect of this correction on pressure turns out to be nearly  
independent of temperature. 

A comparison between the curves corresponding to T$=$ 0 and 50 MeV illustrates the density dependence
of thermal effects. It is apparent that, while providing large to significant contributions at densities  
$\varrho \lesssim 2 \varrho_0$, the thermal pressure plays a negligible role at higher densities, where 
statistical and dynamical effects dominate. 

\subsection{Single-nucleon properties}
\label{sec:QP}

In interacting many-body systems, single-particle states are in general not uniquely defined. Owing to translation 
invariance, however, in nuclear matter they can be identified  using momentum eigenvalues\footnote{For simplicity, here the dependence on discrete quantum numbers is omitted.}, and their properties can be derived within the framework of 
Landau's theory of normal Fermi liquids~\citep{baym-pethick}.
As pointed out  by ~\citet{eos0} for the case of vanishing temperature, the single-nucleon energy of Eq.~\eqref{def:spectrum} 
can be interpreted as the energy of a  {\it quasiparticle} carrying momentum ${\bf k}$ and spin-isospin quantum numbers 
specified by the index $\lambda$.   

\begin{figure*}[ht!]
\includegraphics[scale=0.720]{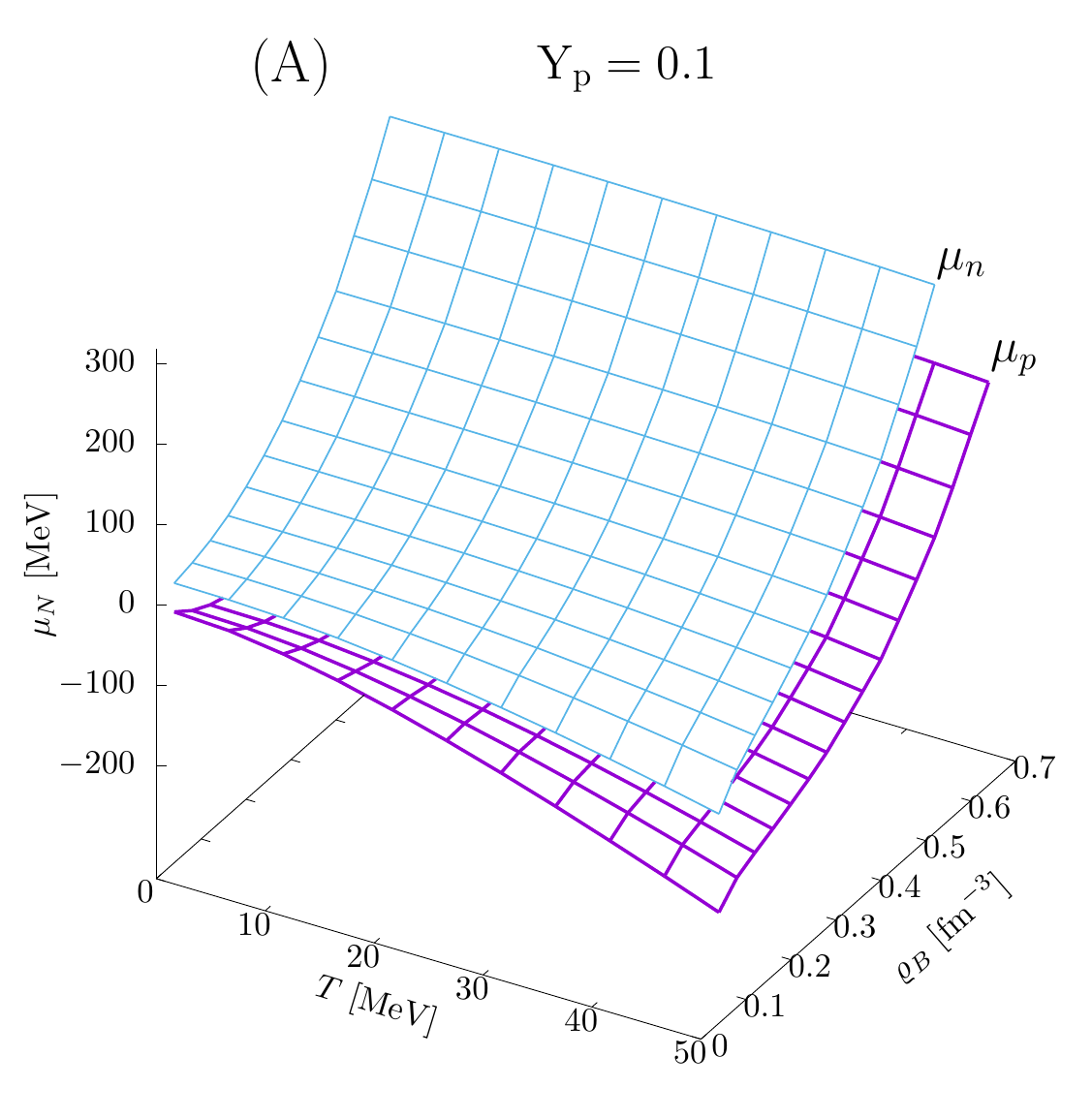}  \includegraphics[scale=0.745]{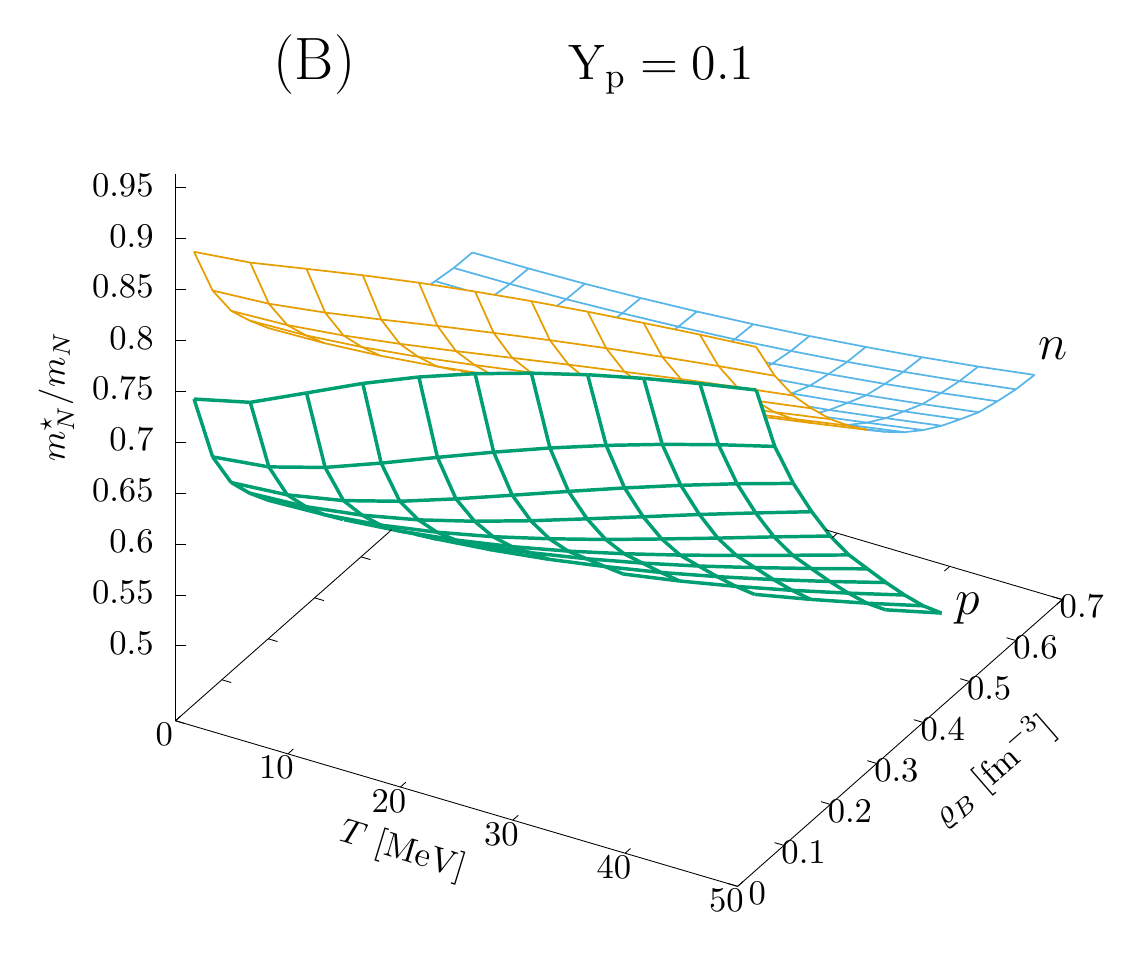}
\caption{(A): density and temperature dependence of the neutron and proton chemical potentials, computed using Eqs.~\eqref{def:E}, \eqref{def:F}, and \eqref{def:S}, with  the CBF effective interaction described in Section~\ref{sec:veff}.  
(B): same as in (A), but for the neutron and proton effective masses, obtained from Eq.~\eqref{def:mstar}, 
given in units of the nucleon mas $m_N$.
 \label{mu_mstar:3D}}
\end{figure*}

The chemical potentials $\mu_\lambda$ are obtained from Eq.~\eqref{det:chempot}. At $T\neq0$, they depend on the spectra 
$e_{k,\lambda}$ through the Fermi distributions, while for $T=0$ they reduce to the Fermi energies $e_{{k_{F_\lambda}}}$, as dictated by  the generalisation of Hugenholtz-Van Hove theorem to the case of a multicomponent system. As an example, in 
the left panel of Fig.~\ref{mu_mstar:3D} the neutron and proton chemical potentials of nuclear matter at proton fraction $Y_p=0.1$\textemdash denoted $\mu_n$ and $\mu_p$, respectively\textemdash are displayed as a function of baryon density and temperature.

The description of quasiparticle dynamics is largely based on the effective mass $m^\star_\lambda$, defined by the equation 
\begin{align}
\label{def:mstar}
\frac{1}{m^\star_\lambda} = \left( \frac{1}{k} \frac{ d e_{k,\lambda} }{d k} \right)_{k = {k_{F_\lambda}}} \ . 
\end{align}
The effective mass, dictating the nucleon dispersion relation in matter, plays a critical role in determining the rates 
of a number of processes relevant to neutron star properties, such as, e.g., neutrino emission and absorption. 

The right panel of Fig.~\ref{mu_mstar:3D} illustrates the density and temperature dependence of the proton and neutron effective masses in nuclear matter at proton fraction $Y_p = 0.1$, expressed in units of the nucleon mass $m_N$. The results show that the 
temperature dependence, often neglected in astrophysical applications, is, in fact, significant. At $\varrho_B = \varrho_0$ 
($3 \varrho_0$) the values of $m^\star_p$ and  $m^\star_n$ at $T=50$ Mev turn out to differ from the corresponding 
zero-temperature values by $\sim 20\%$ and 10\% (15\% and 8\%), respectively.
 
\section{Neutron Star Matter} 
\label{nstar}

The calculation of neutron star properties requires the determination of the EOS of charge neutral
matter consisting of neutrons, protons and leptons in equilibrium with respect to the neutron $\beta$-decay and lepton capture
processes
\begin{align}
n \to p + \ell^- + {\bar \nu}_\ell \ \ \ \ , \ \ \ \ p + \ell^- \to n + \nu_\ell \ ,  
\label{beta:processes}
\end{align}
where $\ell$ denotes the flavour of the lepton participating in the reactions.

For $T=0$, the results of calculations of the EOS, described  in the next section, provide the input needed to solve the 
Tolman-Oppenheimer-Volkoff (TOV) equations~\citep{T,OV}, determining mass and radius of cold neutron stars.  

\subsection{Equation of state of $\beta$-stable matter} 
\label{beta:stability}

Let us consider matter comprising electrons and muons, hereafter referred to as $npe\mu$ matter. Under 
the assumption of transparency to neutrinos and antineutrinos, implying that these particles can freely escape and 
have vanishing chemical potentials, the equilibrium condition reduces to 
\begin{align}
\label{beta}
\mu_n -  \mu_p =  \mu_e  = \mu_\mu \ , 
\end{align}
where $\mu_\ell$ denotes the chemical potential of leptons of flavour $\ell$. The above condition 
must be fulfilled together with the additional constraint of charge neutrality, requiring that 
\begin{align}
\label{charge}
{\rm Y}_p  = {\rm Y}_e + {\rm Y}_\mu \ ,  
\end{align}
with ${\rm Y}_\ell = \varrho_\ell/\varrho_B$, $\varrho_\ell$ being the lepton density. For any given values of 
$\varrho_B$ and $T$, Eqs.~\eqref{beta} and \eqref{charge} univocally determine the proton and lepton fractions, 
needed to obtain the EOS of $\beta$-stable matter.  

\begin{figure}[ht!]
\begin{center}
\includegraphics[scale=0.720]{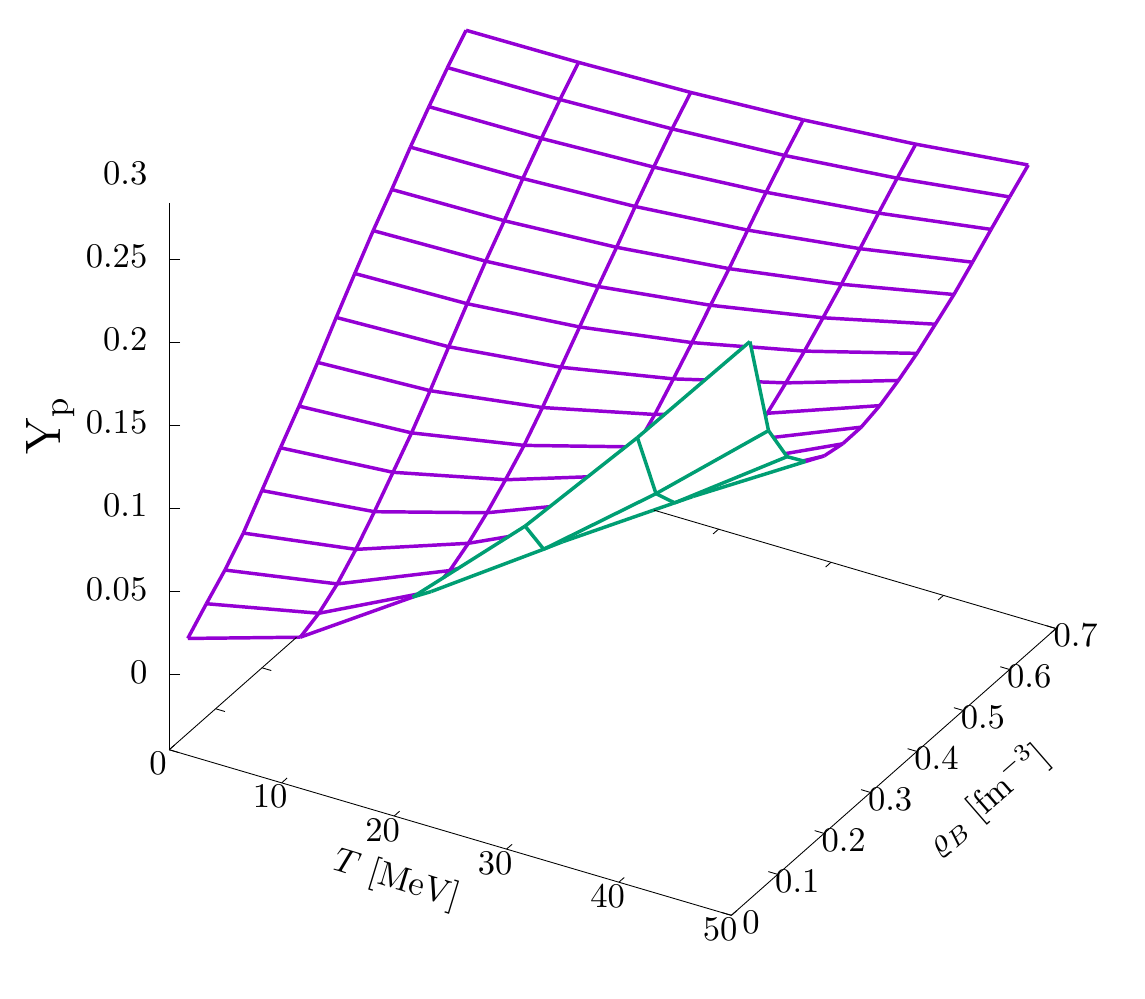}  
\caption{Density and temperature dependence of the proton fraction of charge neutral $\beta$-stable matter, computed using the CBF effective interaction described in Section~\ref{sec:veff}.  
\label{fractions}}
\end{center}
\end{figure}

Figure~\ref{fractions} shows the density and temperature dependence of the proton fraction of charge neutral
$npe\mu$ matter in  $\beta$ equilibrium. The most prominent thermal effect turns out to be a  
departure from the monotonically increasing behavior of $Y_p$ as a function of density, leading to  the appearance of
pronounced minimum for $T \gtrsim 30$~MeV.


In addition to the baryon contributions discussed in the previous sections, the free energy and pressure 
of $\beta$-stable $npe\mu$ matter comprise contributions arising from the presence of leptons, which can be safely treated
as non interacting relativistic fermions. 

\begin{figure}[ht!]
\begin{center}
\includegraphics[scale=0.6750]{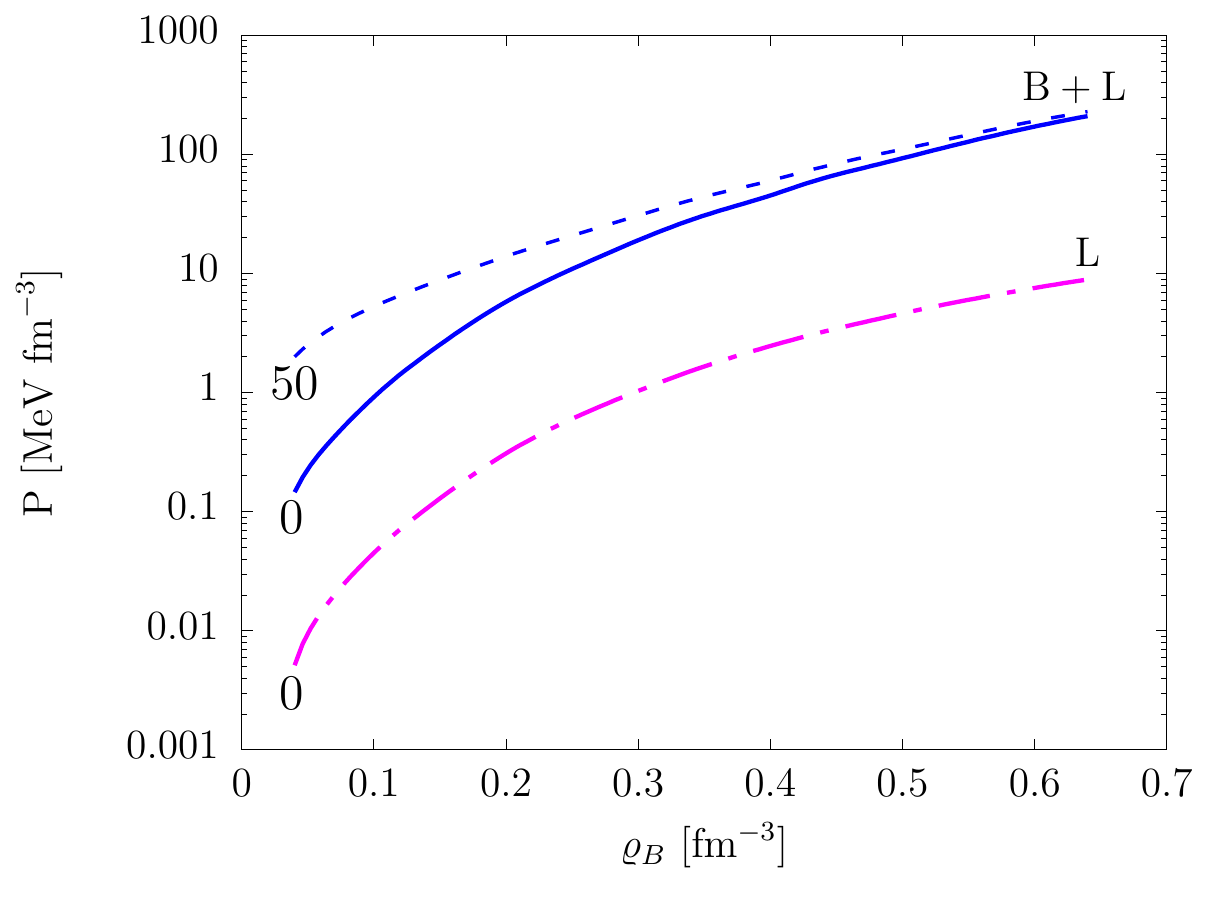}
\end{center}
\caption{Pressure of charge neutral $npe\mu$ matter in  $\beta$-equilibrium, as a function of baryon number density.
The solid and dashed lines show the total pressure of Eq.~\eqref{press:tot} at temperature $T=0$ and 50 MeV,  respectively.
For comparison, the lepton contribution at $T=0$, is also displayed by the dot-dash line. 
 \label{EOS:beta}}
\end{figure}

The solid and dashed lines of Fig.~\ref{EOS:beta} illustrate the density dependence of the total pressure of $npe\mu$ matter
\begin{align}
\label{press:tot}
P(\varrho_B) = P_B(\varrho_B)  + P_L(\varrho_B)  \ , 
\end{align}
where the indices $B$ and $L$ label the baryon and lepton components, 
for $T=0$ and 50 MeV, respectively. A comparison between the
zero-temperature total and lepton pressure, displayed by the dot-dash line, shows that the baryon contribution is largely dominant
at all densities.

\subsection{Neutron star properties} 
\label{MR}

The theoretical framework described in the previous sections has been employed to evaluate a variety of 
properties of cold neutron stars~\citep{sabatucci,bayes1,lucas_QNM}. The results of these analyses show that the predictions of our 
approach are compatible with the determinations of maximum mass, tidal deformability and radius based on recent 
multimessenger observations~\citep{Antoniadis:2013pzd,Cromartie:2019kug,PhysRevLett.119.161101,NICER}.
A pioneering application of the finite-temperature EOSs to study the evolution of proto-neutron stars  
has been also carried out by~\citet{gct}.

\begin{figure}[ht!]
\begin{center}
\includegraphics[scale=0.670]{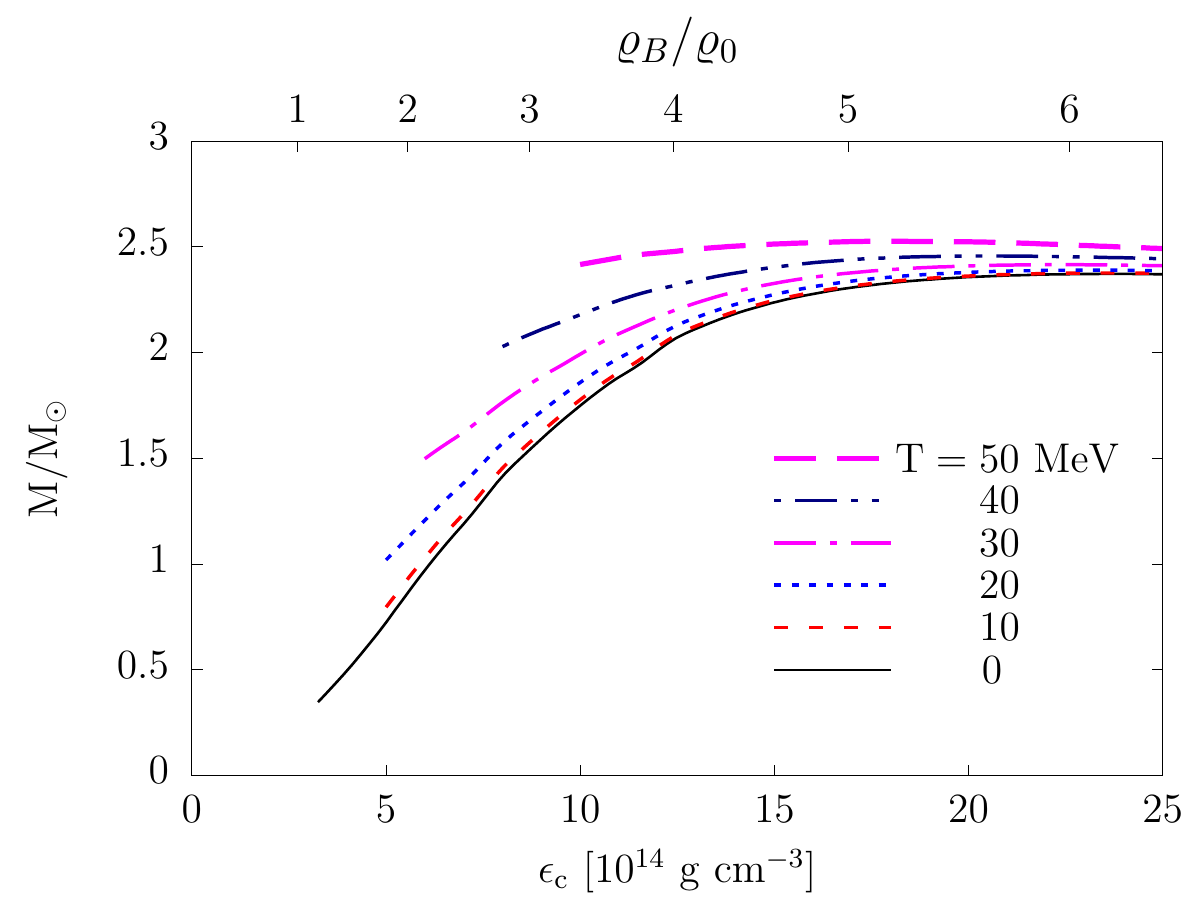}  
\caption{Hot neutron star mass obtained from the TOV equations using the finite-temperature EOSs discussed in the text.
Top and bottom labels correspond to central baryon density\textemdash in units of the saturation density of SNM\textemdash and 
central mass-energy density, respectively.   
\label{max:mass}}
\end{center}
\end{figure}

The determination of the mass-radius relation of hot neutron stars involves non trivial additional difficulties, 
because hydrodynamic equilibrium and heat transfer\textemdash  determining the temperature profile in the star interior\textemdash 
must be consistently taken into account. 
 In order to provide an admittedly rough estimate of the impact of thermal effects on the maximum neutron star mass, 
in Fig.~\ref{max:mass} we show results obtained by solving the TOV equations with EOSs evaluated at constant 
temperature using our approach. 

In the region $\varrho_B/\varrho_0 > 4$, the EOSs computed following the procedure 
discussed in Sect.~\ref{beta:stability} with $0~\leq~T \leq~50$~MeV have been smoothly matched to the zero-temperature 
EOS obtained by~\citet{akmal:1998} using the AV18+UIX nuclear Hamiltonian, which is used as baseline for the 
determination of the CBF effective interaction at $0.25 \leq \varrho_B/\varrho_0 \leq 4$; see Section~\ref{sec:dynamics}. The validity of this prescription is supported by the observation that the contribution of thermal pressure becomes vanishingly small at $\varrho \sim 4 \varrho_0$; see Fig.~\ref{EOS:beta}.

The crust region has been described using the model adopted by~\citet{akmal:1998}, which provides a description of matter at $\varrho_B < 0.1$ fm$^{-3}$.
The use of a zero-temperature EOS appears to be justified for the purpose of our exploratory analysis, because the value of the maximum mass is 
largely unaffected by the structure of matter in the low-density crust region. 
 
The results of Fig.~\ref{max:mass} show that the impact of thermal effects on the maximum mass $M_{\rm max}$, leading to its increase, is not large. 
The values of $M_{\rm max}$ at $T=0$ and 50 MeV turn out to be 2.37 $M_\odot$ and 2.49 $M_\odot$, with  $M_\odot$ being the solar mass, respectively.

It has to be pointed out that the pattern emerging from our analysis\textemdash while being in qualitative agreement with the results of 
existing studies carried out within the framework of the relativistic mean field (RMF) approach or using Skyrme-type models~\citep{Prakash:1997,Kaplan:2014}\textemdash 
turns out to be at variance with
the findings of studies based on G-matrix perturbation theory and phenomenological nuclear Hamiltonians, which predict a 
small decrease of the maximum mass at nonzero temperature~\citep{figuraetal2020,figuraetal2021}.

\section{Summary and outlook} 
\label{summary}

We have developed a consistent formalism suitable to carry out perturbative calculations
of the properties of both cold and hot nuclear matter at arbitrary proton fraction. The distinctive feature of our 
method is the use of an effective nuclear Hamiltonian, based on a phenomenological 
potential providing an accurate description of  nucleon-nucleon scattering in the energy range relevant  to
dense matter in the neutron star interior.  

Our results reproduce the zero-temperature EOSs of PNM and SNM obtained from highly refined 
many-body approaches, such as AFDMC and FHNC/SOC, by construction. 
Moreover, SNM calculations performed using the CBF effective interaction yield the correct equilibrium density, 
and an interaction energy within $\sim15$\% of the empirical value~\citep{eos0}. 

Recently,~\citet{Lovato:2022}, have carried out extensive analysis of the 
EOS of  zero-temperature PNM, aimed at comparing different nuclear Hamiltonians and computational techniques. 
The results of this study, showing that the AV6P+UIX Hamiltonian yields results in close agreement 
with those obtained from the full AV18+UIX up to $\varrho_B = 2 \varrho_0$, strongly supports
the use of this somewhat simplified model to describe nuclear dynamics at supranuclear density.

Within our computational scheme, based on finite-temperature perturbation theory, thermal effects on the 
energy per baryon and the single-nucleon spectrum are taken into account in a consistent fashion, without introducing any simplifying 
assumptions, and the thermodynamic consistency condition turns out to be fulfilled to very high accuracy. 

The results reported in this work suggest that our approach has the potential  to provide accurate evaluations of the 
variety of  properties of dense nuclear matter needed for numerical simulations of astrophysical processes. 

It is very important to note that the availability of a reliable description of neutron star matter in the regime 
in which nucleons are the relevant degrees of freedom is also essential to firmly establish the occurrence of transitions 
to more exotic forms of matter\textemdash involving baryons other than protons and neutrons as well as deconfined quarks\textemdash 
which are expected to become energetically favoured at higher densities. 

The analyses of recent astrophysical data have provided information about the radius of a neutron star of mass $M=1.4 \ M_\odot$, whose central density
does not exceed ~$\sim 3\varrho_0$. The reported values\textemdash $R = 12.45 \pm 0.65$~km~\citep{Miller_2021}, 
12.33$^{+0.76}_{-0.81}$ km~\citep{NICER}, 
and 12.18$^{+0.56}_{-0.79}$ km~\citep{Raaijmakers_2021}\textemdash turn out to be compatible with the predictions of theoretical calculations 
performed using EOSs of purely nucleonic matter~\citep{sabatucci}, thus suggesting that in this case the paradigm of nuclear many-body theory can still be applied.  

The equatorial radius of the neutron star J0740+6620, having mass $M= {2.072}_{-0.066}^{+0.067} \  M_\odot$ and central density   
$\sim 4 \varrho_0$,
 has been also evaluated 
to be  $R = {12.39}_{-0.98}^{+1.30}$~km~\citep{NICER}.  The small difference between the radii of stars 
of mass 1.4 and 2.071 $M_\odot$, implying that the EOS is still rather stiff at $\varrho > 3\varrho_0$, appears to rule out 
the occurrence of a strong first order phase transition in the density range $3 \varrho_0 \lesssim \varrho_B \lesssim 4 \varrho_0$.

Independent empirical information about the limits of applicability of the description of dense nuclear matter in terms of nucleons
is obtained from the analysis of the large body of electron-nucleus scattering data.  
Scaling in the variable $y$\textemdash the definition and physical interpretation of which is thoroughly 
discussed by~\citet{yscaling:west} and~\citet{yscaling}\textemdash has been observed by experiments using a broad range of targets, extending from 
\isotope[2][]{H}~\citep{yscaling_Ciofi} to nuclei
as heavy as \isotope[197][]{Au}~\citep{yscaling_Day,yscaling_Arrington}. The results of these studies unambiguously demonstrate that in the kinematical region corresponding to momentum transfer 
$q \gtrsim 1$~GeV and large negative $y$ the beam particles primarily couple to nucleons belonging to correlated pairs,  
the momentum of which  can be in excess of 500~MeV. Based on the results of experimental analyses carried out at 
Jefferson Lab,~\citet{Subedi_2008}, argued that the occurrence of strong nucleon-nucleon correlations is associated to large fluctuations 
of the nuclear density, that can locally reach values as high as $\sim 5 \varrho_0$. According to the argument proposed in Section~\ref{sec:ham}, this estimate is 
consistent with the observation of scaling at $y \sim -600$~MeV, which in turn implies that at $\varrho_B \sim 5 \varrho_0$ nuclear matter largely behaves as a collection of 
nucleons.

As a final remark, it has to be pointed out that  the description of nuclear matter based on non relativistic nuclear many-body theory, providing the conceptual 
framework underlying our approach, unavoidably leads to violation of causality in the high-density limit.  It has to be pointed out, however, that within the approach 
discussed in this paper the speed of  sound in cold 
$\beta$-stable matter becomes larger than the speed of light at $\varrho > 4.5 \varrho_0$. Therefore, our EOS  turns out to be relativistically consistent in the region relevant to the stability of a $2.2 \ M_\odot$ neutron star; see Fig.~\ref{max:mass}. The difficulties arising from violation of causality can 
be further alleviated taking into account relativistic boost corrections to the nucleon-nucleon potential, along the line discussed by~\citet{akmal:1998}.
Work is presently being carried out to include these correction in our formalism.




\subsection*{Acknowledgements}
The present research is supported by the U.S. Department of Energy, Office of Science, Office of Nuclear Physics, under contracts DE-AC02-06CH11357 (AL).
OB gratefully acknowledges support from the Italian National Institute for Nuclear Research (INFN), under  grant TEONGRAV, as well as the hospitality of the 
CERN Department of Theoretical Physics, where this work was completed. 
GC acknowledges support from the Polish National Research Centre (NCN) grant OPUS 2019/33/B/ST9/00942.

\clearpage


\end{document}